\documentclass[10pt,twocolumn,superscriptaddress,floats,showpacs,nobalancelastpage,longbibliography]{revtex4-1}

\usepackage{array,longtable}
\newcolumntype{L}{>{\tiny $}p{0.33\columnwidth}<{$}}
\newcolumntype{M}{>{\scriptsize $}p{0.33\columnwidth}<{$}}
\newcolumntype{N}{>{\scriptsize $}p{0.43\columnwidth}<{$}}
\usepackage{amsmath}
\usepackage{amssymb}
\usepackage{amsfonts}
\usepackage{graphicx}
\usepackage{hyperref}

\usepackage{float}
\usepackage{graphics}
\usepackage[caption=false]{subfig}

\allowdisplaybreaks[1]

\newif\ifhyper
\hypertrue
\ifhyper
\hypersetup{
   citecolor = {green},
   colorlinks = {true}, 
   urlcolor = {blue} 
}
\fi

\begin{document}

\title{Non-Abelian chiral spin liquid in a quantum antiferromagnet \\ revealed by an iPEPS study}

\author{Ji-Yao Chen}
\affiliation{Laboratoire de Physique Th\'eorique, IRSAMC, Universit\'e de Toulouse, CNRS, UPS, France}
\author{Laurens Vanderstraeten}
\affiliation{Department of Physics and Astronomy, University of Ghent, Krijgslaan 281, 9000 Gent, Belgium} 
\author{Sylvain Capponi}
\affiliation{Laboratoire de Physique Th\'eorique, IRSAMC, Universit\'e de Toulouse, CNRS, UPS, France}
\author{Didier Poilblanc}
\affiliation{Laboratoire de Physique Th\'eorique, IRSAMC, Universit\'e de Toulouse, CNRS, UPS, France}

\date{\today}

\begin{abstract}
Abelian and non-Abelian topological phases exhibiting protected chiral edge modes are ubiquitous in the realm of the
Fractional Quantum Hall (FQH) effect. Here, we investigate a spin-1 Hamiltonian on the square lattice which could, potentially, host the spin liquid analog of the (bosonic) non-Abelian Moore-Read FQH state, as suggested by Exact Diagonalisation
of small clusters. Using families of fully SU(2)-spin 
symmetric and translationally invariant chiral Projected Entangled Pair States (PEPS), variational energy optimization is performed using infinite-PEPS methods, providing good agreement with Density Matrix Renormalisation Group (DMRG) results. A careful analysis of the bulk spin-spin and dimer-dimer correlation functions in the optimized spin liquid suggests that they exhibit long-range ``gossamer tails''. We argue these tails are finite-$D$ artifacts of the chiral PEPS, which become irrelevant when the PEPS bond dimension $D$ is increased.
From the investigation of the entanglement spectrum, we observe sharply defined
chiral edge modes following the prediction of the SU(2)$_2$ Wess-Zumino-Witten theory and exhibiting a conformal field theory (CFT) central charge $c=3/2$, as expected for a Moore-Read chiral spin liquid. 
We conclude that the PEPS formalism offers an unbiased and efficient method to investigate non-Abelian chiral spin liquids in quantum antiferromagnets. 

\end{abstract}
\maketitle

\section{Introduction and model}

The two-dimensional (2D) electron gas experiencing long-range Coulomb repulsion and subject to a strong magnetic field
-- hence breaking time-reversal (TR) symmetry -- can exhibit plethora of topological Fractional Quantum Hall (FQH) phases
at simple rational filling fractions $\nu$~\cite{Stormer1983}. FQH states are characterized by topological order -- the ground state (GS) degeneracy depends on the system topology~\cite{Wen1990,Wen2013} 
-- and by chiral edge modes localized at the system boundaries (if any) and propagating in one direction only~\cite{Wen1991a,Wen1992}.
Such edge modes are gapless and described by known ($1+1$)-dimensional Conformal Field Theories (CFT). The bulk excitations of the FQH states are fractionalized anyons~\cite{Halperin1984} which could have either Abelian statistics, as in
the Laughlin state~\cite{Laughlin1983}, or non-Abelian statistics~\cite{Wen1991b,Nayak2008}, as in the Moore-Read (MR) state~\cite{Moore1991}. Non-Abelian SU(2)$_k$ anyons (for $k>1$) are described by well-known deformations of SU(2), in which only the first $k+1$ angular momenta $j=0,\frac{1}{2},1,\cdots \frac{k}{2}$ of SU(2) occur.
The MR state harbors $j=\frac{1}{2}$ Ising anyons (realized for $k=2$),
descendants of vortices in (two-dimensional) $p+ip$ superconductors~\cite{Kitaev2001,Alicea2012}, 
and exhibiting simple fusion rules, $\frac{1}{2}\times\frac{1}{2}\rightarrow 0+1$.

Fractional Chern Insulators (FCI)~\cite{Repellin2014,Maciejko2015} 
offer the most direct implementation of the FQH physics on the lattice,
still requiring a (gauge) magnetic field to generate electronic bands with non-trivial topological properties (i.e. non-zero Chern numbers), and strong (local) interactions. 
In the case of {\it Mott} insulators, such as those realizing quantum magnets, the appropriate setting to realize FQH physics is less clear. 
It is well known, nevertheless, since the pioneering work of Kalmeyer and Laughlin (KL)~\cite{Kalmeyer1987}, that simple FQH wavefunctions (such as the Abelian
bosonic $\nu=1/2$ Laughlin state) can be ``localized'' on the sites of a 2D lattice in order to realize chiral (singlet) spin liquids (CSL)~\cite{WWZ1989}, spin analogs of the parent FQH states. However, it is largely unknown whether and under which conditions simple {\it local} Hamiltonians
describing (frustrated) quantum antiferromagnets can host such spin liquids, in particular the non-Abelian ones. Recent numerical investigations of a spin-1/2 chiral Heisenberg antiferromagnetic model (HAFM) on the kagom\'e lattice~\cite{Bauer2014,Wietek2015} suggest that a scalar chiral interaction on all triangular units can indeed stabilize a spin liquid
of the $\nu=1/2$ KL type. Similar Abelian CSL were also uncovered in spin-1/2 chiral antiferromagnets on the triangular lattice~\cite{Wietek2017,Gong2017}. Interestingly, the CSL can also emerge in spin-1/2 time-reversal invariant frustrated magnets~\cite{He2014a,Wietek2015}.
Kitaev's {\it anisotropic} honeycomb model in the presence of an external magnetic field~\cite{Kitaev2006} is, so far, the only indisputable example of a local (lattice) Hamiltonian hosting a {\it non-Abelian} CSL, but local spin-1 Hamiltonians on triangular and 
kagome lattices have also been proposed~\cite{Greiter2009,Liu2018}. 
A definite identification of {\it local SU(2)-invariant} models realizing non-Abelian CSL
is therefore needed and the goal of this study.

Further progress in the field of chiral SL have been launched by the constructions of parent quantum spin Hamiltonians~\cite{Schroeter2007,Thomale2009,Nielsen2012,Greiter2014} designed to host various spin analogs of the FQH liquids. For example, by rewriting KL-like states as
correlators of CFT primary fields, a systematic construction of parent Hamiltonians can be obtained. It turns out that, generically, the obtained parent Hamiltonians show long range (algebraic) interactions.  For example, SU(2)-invariant 
spin-1/2 and spin-1 Hamiltonians with long range 3 site-interactions (like ${\bf S}_i\cdot({\bf S}_j\times {\bf S}_k)$) 
have been found to realize the bosonic (Abelian) Laughlin and (non-Abelian) Moore-Read  FQH 
phases, respectively~\cite{Nielsen2012,Glasser2015}. Using such a construction, it was also shown that the spin-1/2 KL spin liquid exhibits the expected chiral edge states~\cite{Herwerth2015}.
Furthermore, it was argued that {\it local} chiral antiferromagnetic Heisenberg models based on some truncation and fine-tuning of the parent Hamiltonians also host the same topological Abelian~\cite{Nielsen2013,Nielsen2014} and non-Abelian phases~\cite{Glasser2015}. 
Similarly to the non-Abelian Kitaev's phase on the hexagonal lattice~\cite{Kitaev2006,Zhu2018}, the spin-1 non-Abelian CSL is expected to host Ising anyons in the bulk. 
However, the proposed spin-1 local chiral HAFM is quite far from the initial parent Hamiltonian and
its detailed investigation is called for.

Besides KL and CFT constructions, topological chiral spin liquids can also be designed using the framework of Projected Entangled Pair States (PEPS)~\cite{Cirac2009b,Cirac2012a,Orus2013,Schuch2013b,Orus2014}, a class of 2D tensor networks~\cite{Nishino2001}. Generally, topological order can be easily implemented in PEPS from local gauge symmetries~\cite{Schuch2013a}. The simplest chiral PEPS is based on a chiral extension~\cite{Poilblanc2015,Poilblanc2016} 
of the spin-1/2 Resonating Valence Bond (RVB) state~\cite{Tang2011,Poilblanc2012,Schuch2012}, originally defined by Anderson as an equal-weight superposition of valence bond configurations~\cite{Anderson1973}. Such a simple PEPS turned out to be critical although, surprisingly, exhibiting 
well-defined chiral edge modes
consistent with the SU(2)$_1$ Wess-Zumino-Witten (WZW) CFT of central charge $c=1$.  A more general and systematic construction of PEPS chiral (and non-chiral~\cite{Poilblanc2017a}) spin liquids has been made recently possible, thanks to 
a general classification 
of SU(2) and translationally invariant PEPS according to their symmetry properties under point group operations~\cite{Mambrini2016}. 
Combining this classification with a Corner Transfer Matrix Renormalization Group (CTMRG) algorithm~\cite{Nishino1996}, one of us investigated the physics of the simple spin-1/2 chiral HAFM mentioned above~\cite{Poilblanc2017b}. Topological order was identified from sharply defined chiral edge modes but,
surprisingly, numerical results suggested critical correlations in the bulk, as for the simpler chiral RVB analog.
Whether this feature is a generic property of chiral PEPS~\cite{Dubail2015} is not clear so far. 
Investigation of new chiral HAFM using PEPS methods is therefore necessary. 

Here, we shall consider the spin-$1$ chiral HAFM defined on the two-dimensional square lattice, as introduced in Ref.~\onlinecite{Glasser2015}~:
\begin{align}
H=J_1 \sum_{\big< i,j\big>} &{\bf S}_i\cdot{\bf S}_j
+ J_2 \sum_{\big<\big<k,l\big>\big>} {\bf S}_k\cdot{\bf S}_l\nonumber \\
+ K_1 \sum_{\big< i,j\big>} &({\bf S}_i\cdot{\bf S}_j)^2 
+ K_2 \sum_{\big<\big< k,l\big>\big>} ({\bf S}_k\cdot{\bf S}_l)^2
\nonumber \\
+ K_c \sum_{\square}\,\, &[{\bf S}_i\cdot({\bf S}_j\times {\bf S}_k)+{\bf S}_j\cdot({\bf S}_k\times {\bf S}_m)\nonumber \\
+ &{\bf S}_i\cdot({\bf S}_j\times {\bf S}_m)+{\bf S}_i\cdot({\bf S}_k\times {\bf S}_m)],
\label{eq:model}
\end{align}
where the first and third sums are taken over nearest-neighbor (NN) bonds and the second and fourth sums run over next-nearest-neighbor (NNN) bonds. The chiral term of amplitude $K_c$ is defined on every 
plaquette of four sites  $(i,j,k,m)$ ordered in (let say) clockwise direction.
The parameters entering (\ref{eq:model}) have been obtained by a careful fine tuning, optimizing the overlap of the exact
GS on small finite-size clusters with the lattice CFT correlator describing the non-Abelian Moore-Read FQH state (on the lattice)~\cite{Glasser2015}. We will here adopt these fine-tuned parameters (retaining only 3 digits),  $J_1=1$, $J_2=0.623$, $K_1=-0.176$, $K_2=0.323$ and $K_c=0.464$. Note that the related spin-1/2 chiral HAFM introduced in Ref.~\onlinecite{Nielsen2012} and studied in Ref.~\onlinecite{Poilblanc2017b} contains only the $J_1$ and $J_2$ bilinear terms and the $K_c$ chiral term 
since the biquadratic interactions $K_1$ and $K_2$ become irrelevant for spin-1/2.

In order to explore the physics of the above model, we combine different numerical techniques such as
Lanczos Exact Diagonalisations (LED),
Density Matrix Renormalization Group (DMRG)~\cite{White1992} and 
tensor network methods~\cite{Cirac2009b,Cirac2012a,Orus2013,Schuch2013b,Orus2014}, all reviewed in 
Sec.~\ref{sec:methods}.
In particular, we shall focus on \hbox{spin-1} SU(2)-symmetric PEPS to describe the chiral spin liquid phase. 
More precisely, we construct (disconnected) families of PEPS breaking time-reversal (T) and parity (P) symmetries -- without breaking PT -- providing a faithful representation of chiral spin liquids directly in the thermodynamic limit. 
In contrast to usual PEPS calculations, which approach the ground state of the model via imaginary time evolution (and could get trapped in local minima), we use a more elegant and secure framework based on a variational optimization scheme (combined with a CTMRG algorithm), taking advantage of the reduced number of variational parameters in the fully symmetric ansatz. 

Using such state-of-the-art numerical techniques, we shall address a number of important issues. 
First, in Sec.~\ref{sec:bulk},  we shall investigate the property of the bulk system, i.e. whether it exhibits 
short range correlations like its "parent" FQH Moore-Read state or whether it is critical such as the spin-1/2 chiral PEPS analog. Second, in Sec.~\ref{sec:edge}, we shall consider the edge spectrum, seeking to characterize topological chiral order, and looking for evidence of its non-Abelian character. 
Finally, we shall discuss the results in the last section, and make some conjecture.  Experimental setups will also be briefly discussed. 

\section{Short summary of numerical methods}
\label{sec:methods}

\subsection{Lanczos exact diagonalisations}

In Refs.~\onlinecite{Glasser2015}, the parameters of the spin-1 model have been obtained using exact diagonalization 
of small lattices (up to $4\times 4$) on the plane or on the cylinder, 
fine tuning the overlap of the exact GS with the targeted non-Abelian chiral state. 
Here we diagonalize, using Lanczos ED methods,  $4\times 4$ and $\sqrt{20}\times\sqrt{20}$ square tori -- exhibiting the full translation and (at least) $C_4$ point group symmetries of the square lattice -- to investigate the low-energy spectrum of the model. In contrast to the planar geometry, in the case of a torus geometry the GS  is expected to become degenerate (3-fold degenerate for the Moore-Read
state) in a gapped topological phase and in the limit of very large sizes. Hence, fundamental differences from the previous computations are expected, even on small clusters. 

\subsection{Density Matrix Renormalization Group}

A standard approach for matrix product state (MPS) simulations of 2D systems consists in studying cylinders (with $N=L\times W$ sites, $L>W$) with periodic (respectively open) boundary conditions in the short (respectively long) dimension. We have used the ITensor library for these calculations~\footnote{ITensor C++ library, available at \url{http://itensor.org}.}, in particular using total $S_z$ conservation. 
We have observed that convergence is very hard to achieve for $W>8$ due to the large entanglement entropy of a half-system and the maximum number of states kept in the simulation was $m=2,000$. This may be related to criticality of the system (see later) or large correlation length. Therefore, we have used MPS computations as a way to extract the ground-state energy per site $e_0(L,W)$. First, for a given system size $L\times W$, we can extrapolate the total energy vs discarded weight. Then, at fixed width $W$, we can estimate $e_0(W)$ by extrapolating $e_0(L,W)$ vs $1/L$. Another estimate of the same quantity can be obtained from different cylinders of width $W$ and lengths $L_1$ and $L_2$, simply by subtracting total energies $E_0$:
\begin{equation}\label{eq:e0_dmrg}
e_0(W) = \frac{E_0(L_2,W)-E_0(L_1,W)}{L_2-L_1},
\end{equation}
which has the advantage to reduce finite-size effects due to the open boundaries on the edges. 
At last, we can extrapolate $e_0(W)$ vs $1/W$ as shown in 
subsection~\ref{subsec:ener}.

Quite interestingly, we can also obtain an \emph{exact upper bound} using cylinders with open boundary conditions (OBC) in both directions, and using their ground-state wavefunctions as a variational one for the infinite system~\cite{Yan2011}. These values are also reported in subsection \ref{subsec:ener}.

\subsection{iPEPS method}
\subsubsection{Singlet chiral PEPS ansatz}
Our infinite-PEPS (iPEPS) method -- directly in the thermodynamic limit -- relies first on the construction of 
generic PEPS ans\"atze fulfilling all the necessary symmetry properties of the targeted chiral spin liquid. First, for convenience, we apply 
a $\pi$-rotation along the $Y$ spin axis on all sites of one of the two sublattices, which enables to
express the (approximate) GS wave function(s) in terms of a unique site tensor $\mathcal{A}_{\alpha\beta\gamma\delta}^{s}$
(instead of two, one for each sublattice), where the greek indices label the states of the $D$-dimensional {\it virtual} space $\mathcal{V}$ attached to each site in the 
$z=4$ directions of the lattice,
and $s=0,\pm 1$ is the $S_z$ component of the physical spin-$1$. 
A chiral spin liquid bears symmetry properties that greatly constrain the PEPS ansatz.
To construct such an ansatz, we use a general classification of all (translationally-invariant)  SU(2)-symmetric (i.e. invariant under any spin rotation) PEPS proposed recently~\cite{Mambrini2016}, 
in terms of the five irreducible representations (IRREPs) $A_1$, $A_2$, $B_1$, $B_2$ and $E$ of the square lattice $C_{4v}$ point group~\cite{LandauQM}. In this setting, 
the virtual space $\mathcal{V}$ is defined as any direct sum of SU(2) IRREPs.
Since the chiral spin liquid only breaks P and T  but does not break the product PT, 
the simplest adequate PEPS complex tensors have the following forms,
\begin{eqnarray}
\label{eq:pepsa}
\mathcal{A} &=& \mathcal{A}_1 + i\mathcal{A}_2 = \sum_{a=1}^{N_1} \lambda^1_a\mathcal{A}_1^a + i\sum_{b=1}^{N_2} \lambda^2_b\mathcal{A}_2^b,\\
\label{eq:pepsb}
\mathcal{B} &=& \mathcal{B}_1 + i\mathcal{B}_2 = \sum_{a=1}^{M_1} \mu^1_a\mathcal{B}_1^a + i\sum_{b=1}^{M_2} \mu^2_b\mathcal{B}_2^b,
\end{eqnarray} 
graphically shown in Fig.~\ref{fig:CTMRG}(a),
where the real elementary tensors $\mathcal{A}_1^a$ ($\mathcal{B}_1^a$) and $\mathcal{A}_2^b$ 
($\mathcal{B}_2^a$) have the same set of virtual spins and transform according to 
the $A_1$ ($B_1$) and $A_2$ 
($B_2$) IRREP of the point group $C_{4v}$ of the square lattice. $N_1$ and $N_2$
($M_1$ and $M_2$) are the numbers of such elementary tensors in each class, respectively.
$\lambda^1_{a}$ and $\lambda^2_{a}$ ($\mu^1_b$ and
$\mu^2_b$) are arbitrary real coefficients of these tensors. 
Note that, in the atomic orbital language, $\mathcal{A}$ and $\mathcal{B}$ would correspond to $s+ig$ and $d_{x^2-y^2}+id_{xy}$ orbitals, respectively. The PEPS wavefunction is
obtained by the contraction of all site tensors (i.e. by summing all virtual indices on the links). 

\begin{table}[hbt]
\renewcommand\arraystretch{1.5}
\caption{
Numbers of independent SU(2)-symmetric spin-$1$ tensors for the different 
virtual spaces we consider, $D\le 6$. The third (fourth) column gives the number of $\mathcal{A}_1 / \mathcal{B}_1$ ($\mathcal{A}_2 / \mathcal{B}_2$) tensors and 
the last column contains the total numbers $\mathcal{D}_A$ / $\mathcal{D}_B$
of tensors entering Eq.~(\protect\ref{eq:pepsa}) / Eq.~(\protect\ref{eq:pepsb}). $*$ means the states of the PEPS family are all real (i.e. non chiral).
}
\label{TABLE:numbers}
\begin{center}
\begin{tabular}{ccccc}
    \hline 
    \hline
    \quad\quad ${\mathcal{V}}$ \quad\quad & \quad $D$ \quad\quad\quad & $N_1 / M_1$ \qquad\quad & $N_2 / M_2$ \qquad\quad & $\mathcal{D}_A$ / $\mathcal{D}_B$ \\
    \hline
    $\frac{1}{2}\oplus 0$ & 3 \quad\quad & 2 / 2 \qquad\quad & 0 / 1 \qquad\quad & 2$^*$ / 3$^*$ \\
      $\frac{1}{2}\oplus \frac{1}{2}$ & 4 \quad\quad & 6 / 9 \qquad\quad & 4 / 3 \qquad\quad & 
    10 / 12 \\
    $1\oplus 0$ & 4 \quad\quad & 3 / 5 \qquad\quad & 3 / 1 \qquad\quad & 
    6 / 6 \\
  $\frac{1}{2}\oplus \frac{1}{2}\oplus 0$ & 5 \quad\quad & 12 / 13 \qquad\quad & 5 / 6 \qquad\quad & 
    17 / 19 \\
  $1\oplus \frac{1}{2}$ & 5 \quad\quad & 5 / 5 \qquad\quad & 3 / 4 \qquad\quad & 
    8 / 9 \\
    $1\oplus \frac{1}{2}\oplus 0$ & 6 \quad\quad & 13 / 13  \qquad\quad & 8 / 9 \qquad\quad & 
    21 / 22  \\
      \hline
    \hline 
\end{tabular}
\end{center}
\end{table}

The symmetric tensors up to $D\le 6$ have been tabulated in Ref.~\onlinecite{Mambrini2016},
and their numbers $\mathcal{D}_A=N_1+N_2$ and $\mathcal{D}_B=M_1+M_2$, for the most relevant virtual spaces $\mathcal{V}$, 
are listed in Table~\ref{TABLE:numbers}.
For each choice of the virtual space, Eqs.~(\ref{eq:pepsa}) and (\ref{eq:pepsb}) 
then provide two {\it families} of PEPS.
 From the table of characters of the $C_{4v}$ IRREPs, it is clear that $\mathcal{A}\rightarrow\bar{\mathcal{A}}$
and $\mathcal{B}\rightarrow\pm\bar{\mathcal{B}}$ under any of the point group reflexions, 
$R_x$ and $R_y$ along the crystallographic axes,
and $R_{x+y}$ and $R_{x-y}$ along the $\pm 45$-degree directions. Hence, the corresponding 
PEPS wave function transforms into its complex conjugate, which is equivalent (for a singlet state) to the effect of applying time reversal symmetry.  

Our chiral PEPS also exhibit a very important gauge symmetry encoded 
at the level of the local $\mathcal{A}$ and $\mathcal{B}$ tensors.
More precisely, the number of spins 1/2 (or half-integer spins, in general) present in the set of virtual degrees of freedom 
attached to each site is always even. The $\mathbb{Z}_2$ gauge symmetry linked to this parity conservation is at the origin of the topological order sustained by the PEPS~\cite{Schuch2013a}. 

\subsubsection{CTMRG   algorithm}

\begin{figure}[htb]
\centering
	\begin{minipage}[c]{0.12\textwidth}
	\centering
	\subfloat[]{\includegraphics[width=16mm, height=16mm]{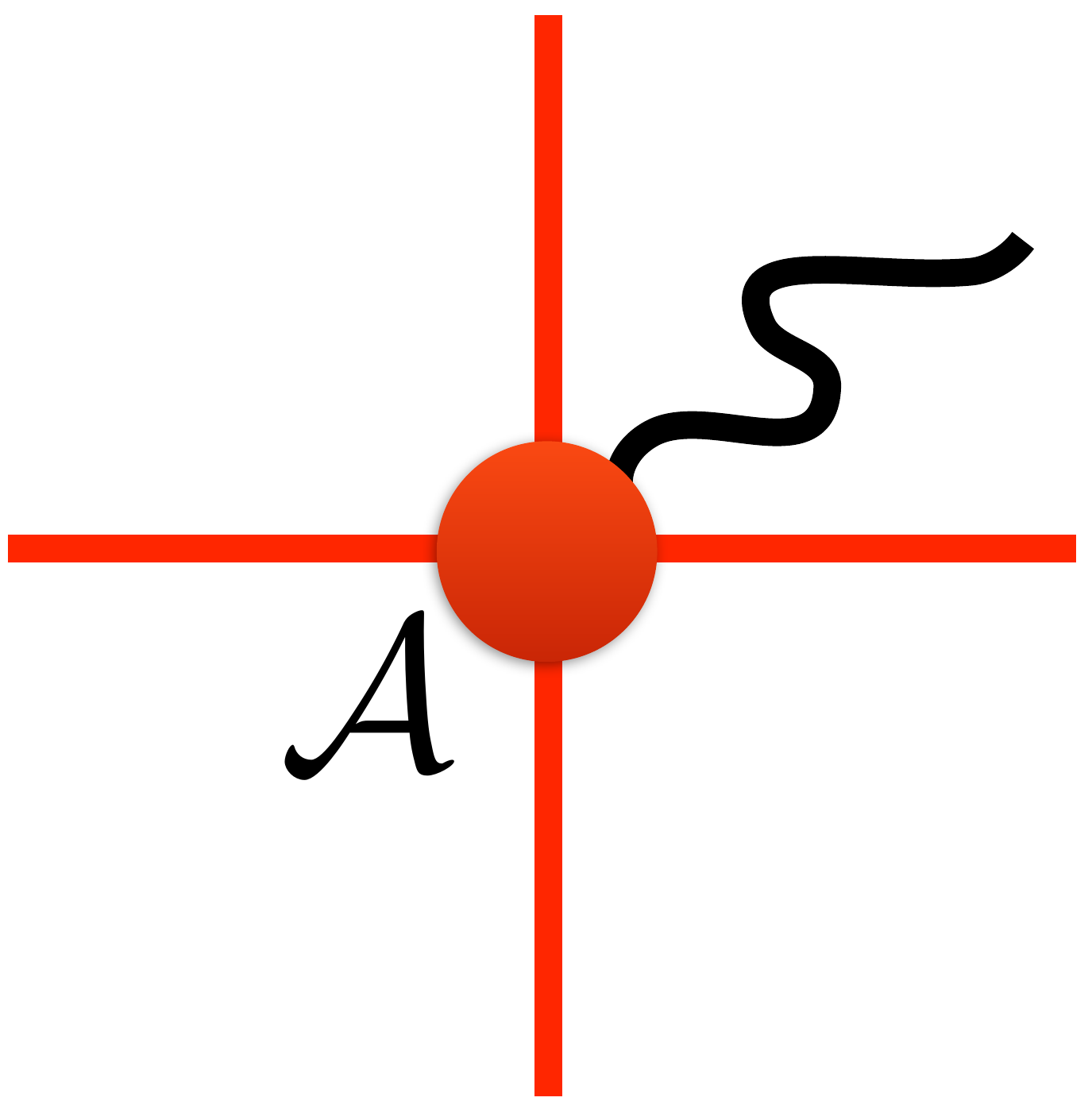}}\\
	\subfloat[]{\includegraphics[width=16mm, height=16mm]{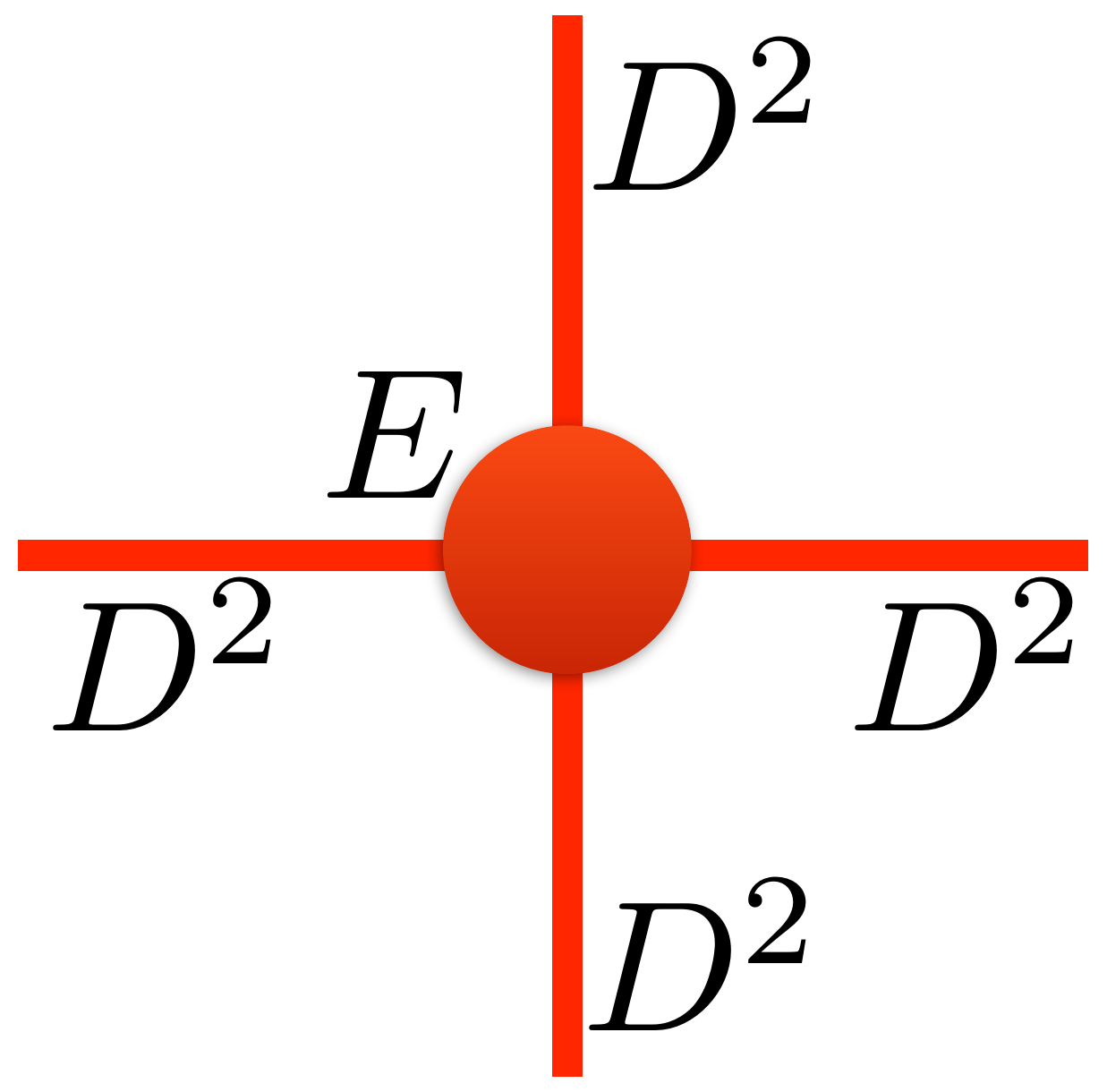}}
	\end{minipage}
	\begin{minipage}[]{0.35\textwidth}
	\centering
	\subfloat[]{\includegraphics[width=43mm, height=41.5mm]{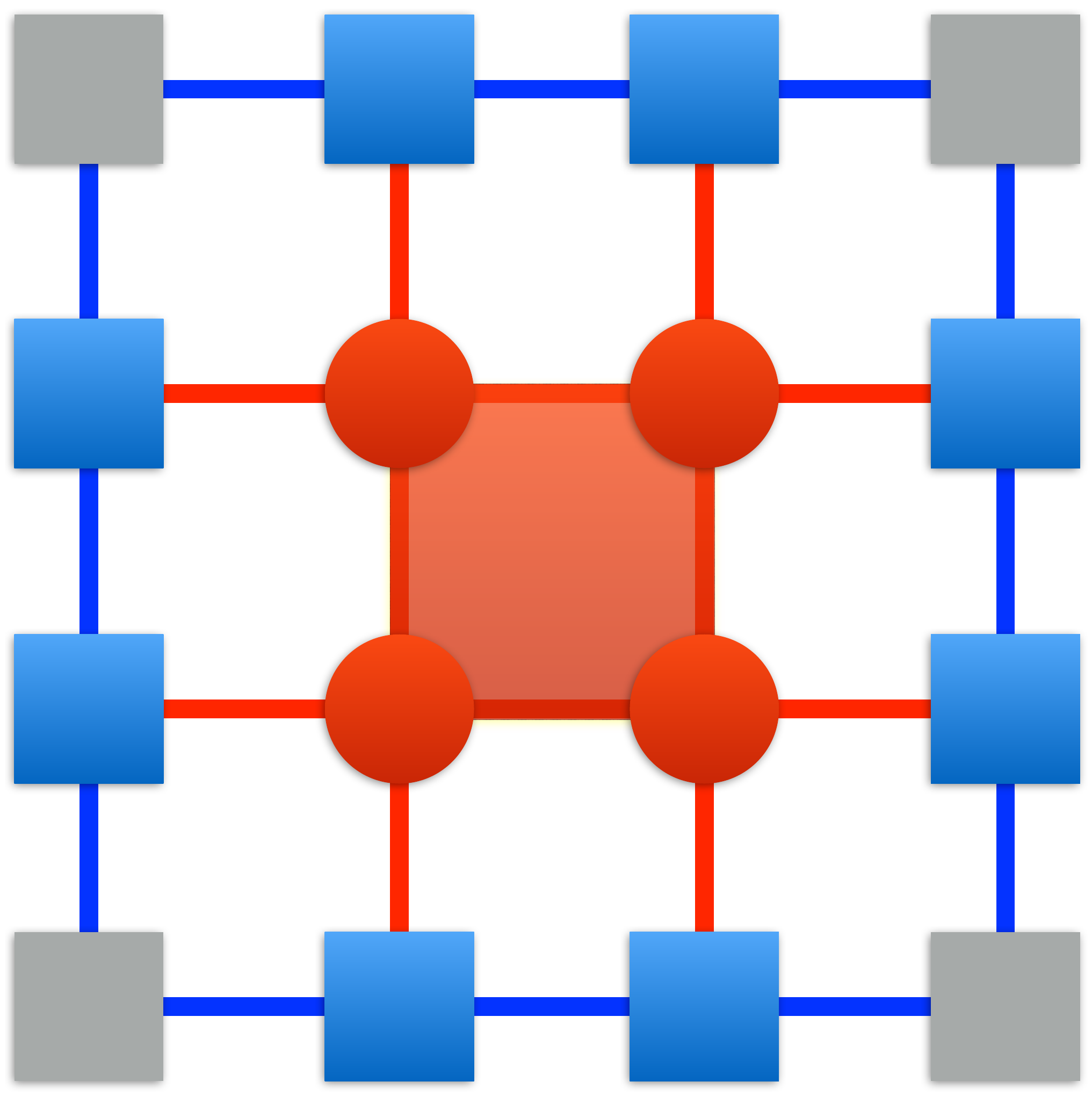}}
	\end{minipage}\\
	\begin{minipage}[c]{0.13\textwidth}
	\centering
	\subfloat[]{\includegraphics[width=16mm, height=16mm]{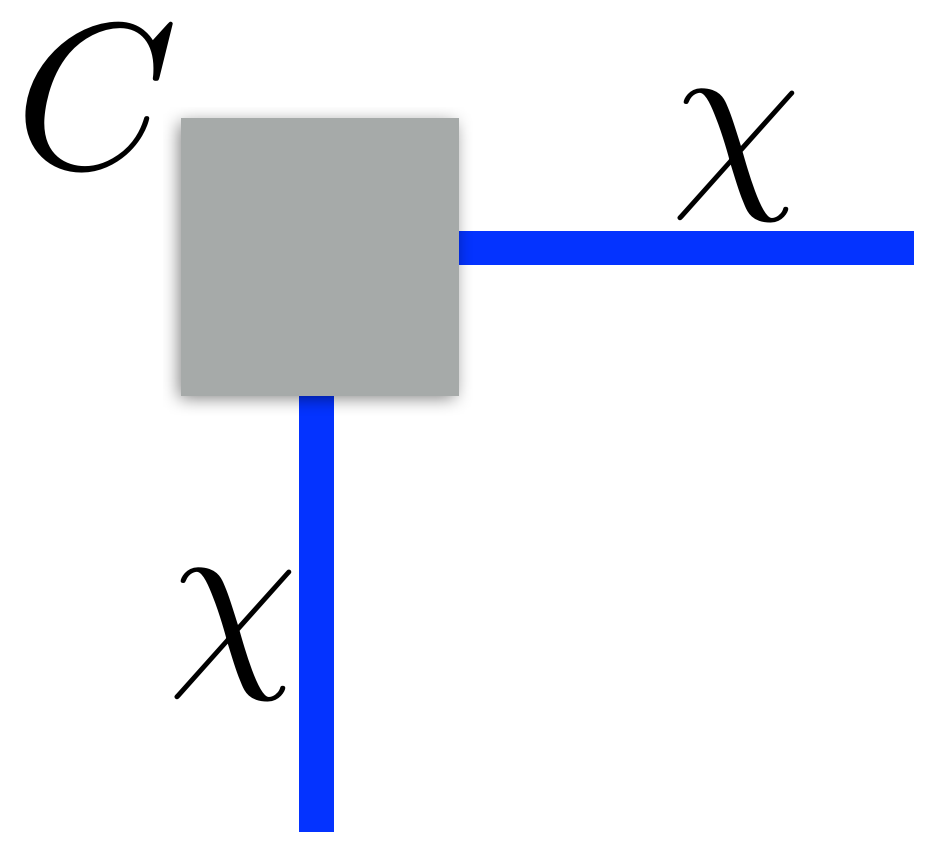}}\\
	\subfloat[]{\includegraphics[width=20mm, height=16mm]{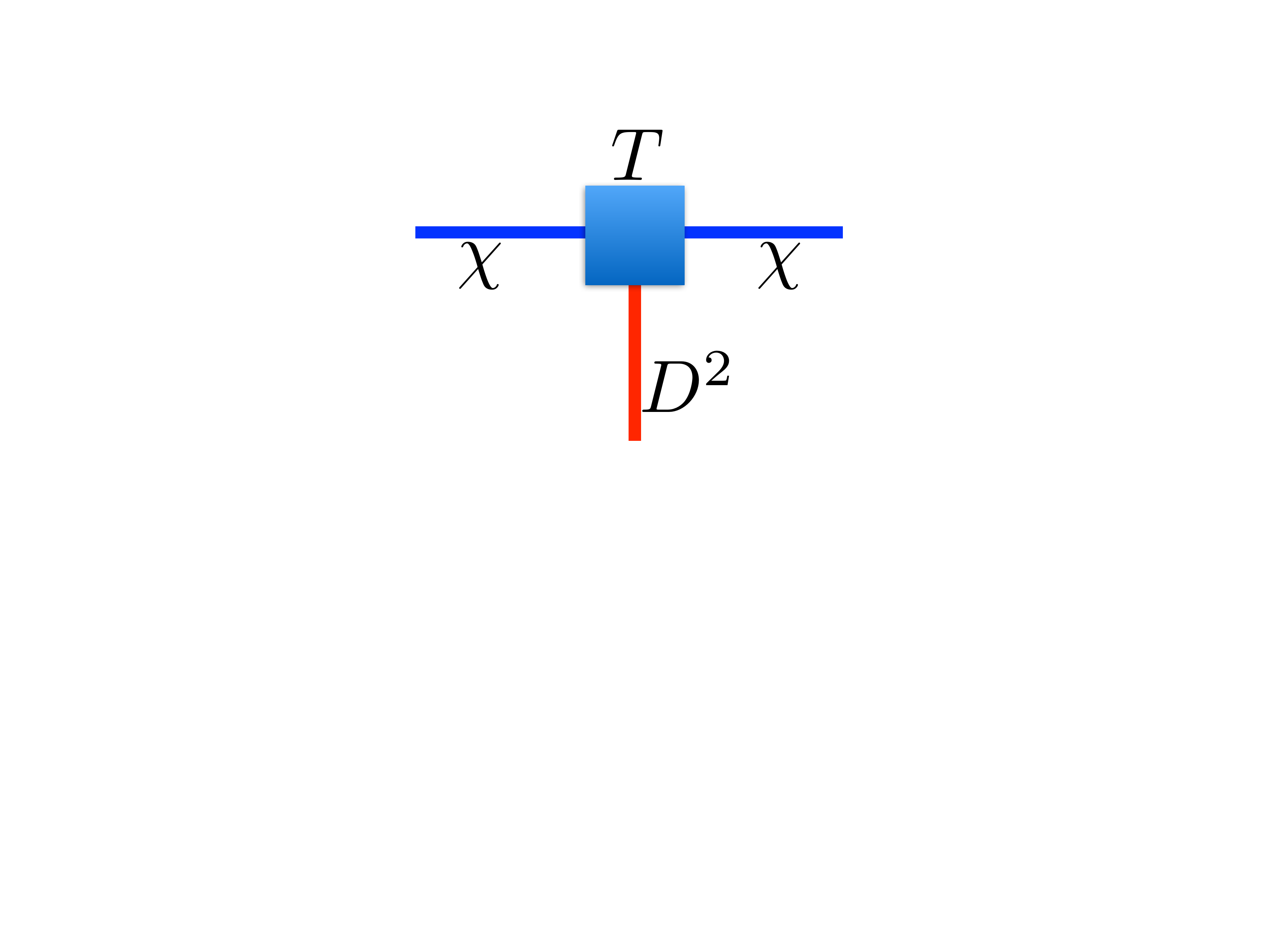}}
	\end{minipage}
	\begin{minipage}[]{0.32\textwidth}
	\centering
	\subfloat[]{\includegraphics[width=24mm, height=26mm]{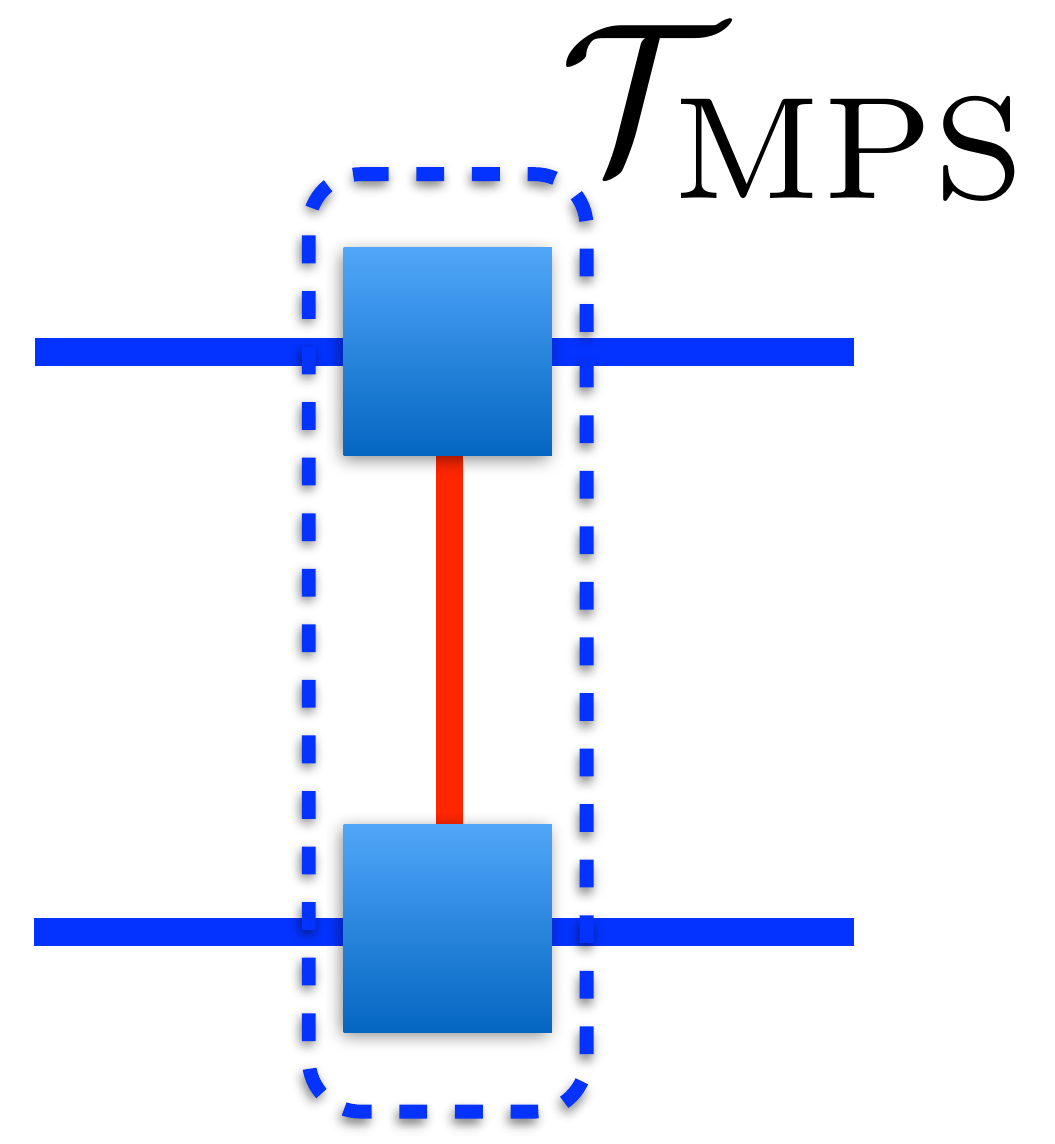}}
	\subfloat[]{\includegraphics[width=20mm, height=34mm]{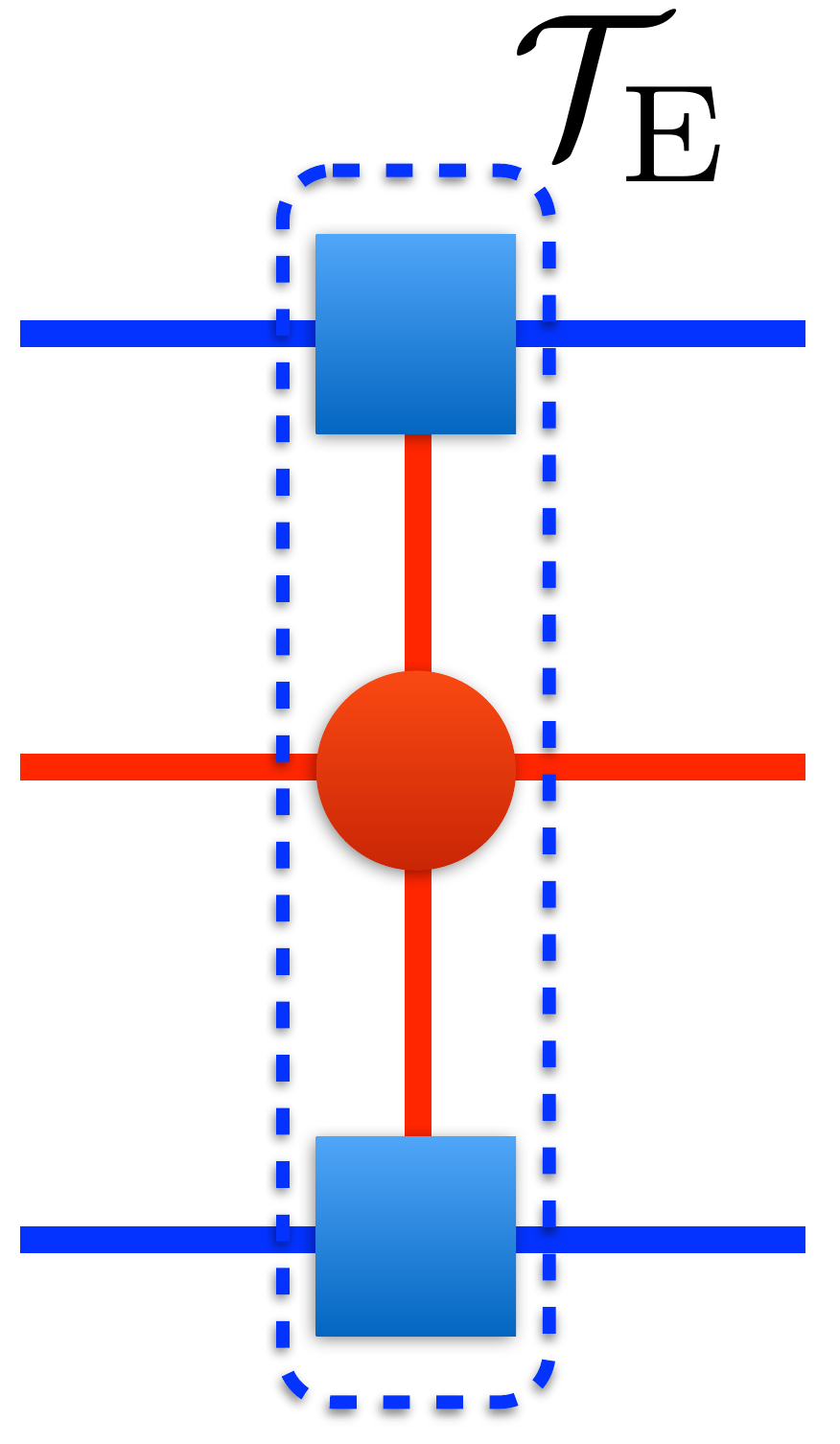}}
	\end{minipage}
\caption{
CTMRG   for the chiral PEPS with one-site unit cell. 
(a) $A_1+iA_2$ PEPS tensor $\mathcal{A}$. (b) Double tensor $E$ obtained by tracing out physical indices $E=\sum_s\bar{\mathcal{A}}^s\mathcal{A}^s$ (or $E=\sum_s\bar{\mathcal{B}}^s\mathcal{B}^s$), 
where $\bar{\mathcal{A}}^s$ ($\bar{\mathcal{B}}^s$) is the
complex conjugate of $\mathcal{A}^s$ ($\mathcal{B}^s$). 
(c) CTMRG   environment for $2\times 2$ cluster, constructed from the (real) corner matrix $C$ and the (hermitian) boundary tensor $T$, depicted separately in (d) and (e), respectively. 
The environment bond dimension is chosen to be $\chi=kD^2 (k\in \mathbb{N}_{+})$. 
The energy density is calculated by inserting either the identity operator or the local hamiltonian operator in the red shaded $2\times 2$ cluster. 
(f) and (g) are the  transfer matrices constructed from the $T$ tensor only, and from the $T$ and $E$ tensors, respectively. 
The maximal correlation length can be obtained from the largest two eigenvalues of 
both transfer matrices (see text).
}
\label{fig:CTMRG}
\end{figure}

Once PEPS families have been constructed, the second step is to optimize the Hamiltonian energy
w.r.t. the tensor parameters, for each class separately. The reduced number of parameters (obtained thanks to the implementation of the full state symmetries) allows to perform a "brute force" optimization.
For each set of PEPS parameters, one then needs 
to compute the corresponding variational energy, in order to "feed" an efficient minimization routine i.e.
one based on a Conjugate Gradient (CG) method. The variational energy computation is done directly
in the thermodynamic limit using the CTMRG
algorithm~\cite{Nishino1996}. After constructing the double-layer tensor $E$ of Fig.~\ref{fig:CTMRG}(b), one obtains, using
a real-space RG method,  the environment of Fig.~\ref{fig:CTMRG}(c) surrounding the active $2\times 2$ region and involving the CTM $C$ and $T$ tensors shown in Figs.~\ref{fig:CTMRG}(d)(e). The Identity matrix or the Hamiltonian is then inserted in the active region (between the two layers) to compute the energy per site. Note that, for our chiral PEPS, all $C$ and $T$ tensors in Fig.~\ref{fig:CTMRG}(c) are identical by symmetry (and the $C$ matrix remains Hermitian after each CTMRG step), 
which simplifies significantly the CTMRG procedure. 

\subsubsection{Uniform MPS method}

An alternative method for computing effective environments for PEPS in the thermodynamic limit relies on uniform matrix product states (MPS). A transfer matrix is constructed by repeating the double-layer tensor $E$ (Fig.1(b)) on an infinite linear chain, and we find the transfer-matrix fixed point as a uniform MPS using variational optimization \cite{Fishman2017}. After repeating this procedure in different lattice directions, we find effective environments and we can compute observables from the PEPS. Additionally, the use of channel environments \cite{Vanderstraeten2015b} allows to compute correlation functions directly in momentum space.

\section{Results on bulk properties}
\label{sec:bulk}
\subsection{Low-energy spectra on small tori}

Let us first investigate the spin-1 chiral HAFM on small 16-site and 20-site clusters with periodic boundary conditions and
full (or partial) point group symmetry, enabling to {\it a priori} block-diagonalize the Hamiltonian matrix according to 
the irreducible representations (IRREPs) of the cluster space group. We also use the total $S_z$ quantum number, enabling to reconstruct the
exact SU(2) multiplet structure of the energy spectrum. 

\begin{figure}[htbp]
\centering
\includegraphics[width=\columnwidth,angle=0]{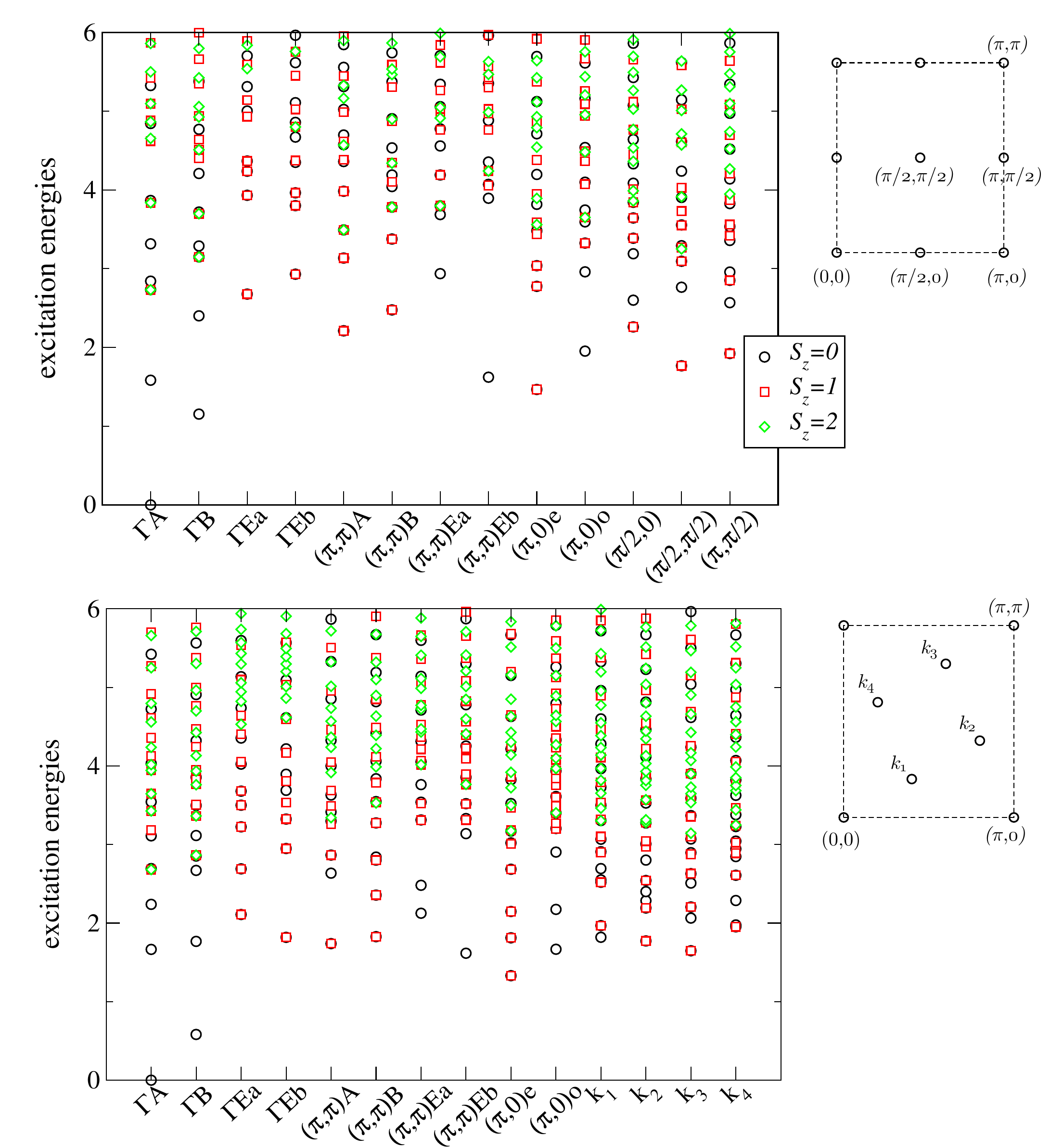}
	\caption{
Lanczos ED of  the spin-1 chiral HAFM on a $4\times 4$ (16-site) (a) and
$\sqrt{20}\times\sqrt{20}$ (20-site) (b) tori. The various columns correspond to different IRREPs
of the space group and different symbols are used to distinguish eigenstates with different (total) spin quantum number. Momenta corresponding to the respective Brillouin zones are shown on the right. The GS energy has been subtracted for clarity. }
\label{fig:LED}
\end{figure}

The low-energy spectra, split in the various IRREPs, are shown in Fig.~\ref{fig:LED}(a,b). For the Moore-Read state, we expect to observe 3 quasi degenerate eigenstates on a torus. In particular, their momentum quantum numbers can be obtained from a simple counting rules~\cite{Regnault2011,Bernevig2012,Laeuchli2013} using partitions $(2,0,2,0,\ldots)$, $(0,2,0,2,\ldots)$ and $(1,1,1,1,\ldots)$, from which we predict that these 3 states should be at the $\Gamma$ point ($K=(0,0)$) for 16- and 20-site square clusters. However, no clear energy gap separating a group of 
quasi-degenerate singlets from the rest of the spectrum -- the signature of the onset of topological GS degeneracy -- is seen. Still, the three lowest singlets are indeed found at the $\Gamma$ point. We believe the cluster sizes are too small compared to some relevant bulk correlation length. 

\subsection{Energy extrapolations}
\label{subsec:ener}

\begin{figure}[htb]
\centering
\includegraphics[width=0.99\columnwidth,angle=0]{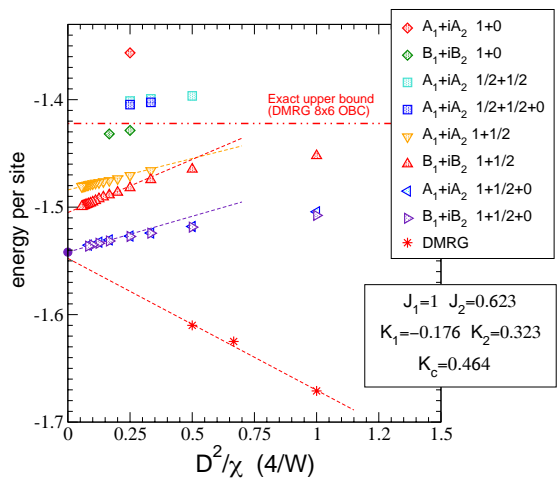}
\caption{
iPEPS variational energies per site versus $D^2/\chi$. Results for $D=5$ and $D=6$ 
have been extrapolated to the $\chi\rightarrow\infty$ limit. 
A comparison to DMRG data obtained on cylinders of width $W$ (see text) and plotted versus $1/W$ ($\times 4$) is shown.
 }
\label{fig:energy_compare}
\end{figure}

Let us first consider the results for the iPEPS energy density plotted in Fig.~\ref{fig:energy_compare} as a function of $D^2/\chi$, for $D=3,4,5$ and various classes corresponding to different virtual spaces. We only restrict here to the classes providing the best energies for a given choice of $D$. 
Note that the variational energy is optimized up to a maximal value of $\chi=\chi_{\rm opt}$ which depends on $D$ (typically, 
$\chi_{\rm opt}=4D^2=100$ for $D=5$ and $\chi_{\rm opt}=3D^2=108$ for $D=6$) 
and then a ``frozen ansatz'' is used for $\chi>\chi_{\rm opt}$. We observe that 
the energy density systematically decreases for increasing $\chi$ so that all data points can be considered as exact 
variational upper bounds.  In fact, the data show a well-behaved scaling, almost perfectly linear in $1/\chi$, so that accurate $\chi\rightarrow\infty$ extrapolations of the energy density can be obtained for each class. The best energies
have been obtained for virtuals spaces $1\oplus\frac{1}{2}$ ($D=5$) and $1\oplus\frac{1}{2}\oplus 0$ ($D=6$).
Note also that $1\oplus 0$ ($D=4$) gives a better energy than $\frac{1}{2}\oplus\frac{1}{2}\otimes 0$ ($D=5$)
so we believe the presence of a spin-1 in the virtual space is essential to get good variational ans\"atze.
Note also that the energy difference between the $\mathcal{A}$ and $\mathcal{B}$ PEPS decreases rapidly with increasing $D$, and become negligible  for $D=6$. We believe the two ans\"atze would describe exactly 
the same state in the $D\rightarrow\infty$ limit.

We then compare the PEPS energies to the DMRG 
results in Fig.~\ref{fig:energy_compare}. The DMRG energy densities computed on cylinders of various widths $W$ (see details above) have been (tentatively) extrapolated in the $W\rightarrow\infty$
limit showing nice agreement with the $D=6$ iPEPS $\chi\rightarrow\infty$ extrapolation. 
This indicates that our symmetric chiral PEPS provide good approximations for the true ground state in the thermodynamic limit, albeit with relatively small bond dimension $D$.

\subsection{Correlation lengths}

\subsubsection{From the TM spectrum}

Correlations lengths can be obtained from the leading eigenvalues of the $\chi^2\times \chi^2$ transfer matrix $\mathcal{T}_{\rm MPS}$ depicted in Fig.\ref{fig:CTMRG}(f). 
Ordering the (real) eigenvalues as $t_0 > |t_1|> |t_2|>\cdots $, one obtains decreasing correlation lengths $\xi^{\rm (n)}_{\rm MPS}$, $n\in \mathbb{N}^+$, defined as
\begin{equation}
\xi^{\rm (n)}_{\rm MPS}=1/\ln{(t_0/|t_n|)}\, .
\label{eq:correl_mps}
\end{equation}
Note that the leading eigenvalue is non-degenerate (and can be normalized to 1)
while the sub-leading eigenvalues may be degenerate. This degeneracy provides
 informations on the type of local operators the correlation length is associated to
 (see later). Alternatively, correlation lengths $\xi_{\rm E}^{(n)}$ can obtained in the same way from the $\chi^2D^2\times D^2\chi^2$ transfer matrix $\mathcal{T}_{\rm E}$, 
 where a $E$ tensor is inserted between
 the two $T$ tensors (see Fig.~\ref{fig:CTMRG}(g)). Since $T$ describes the exact environment
 fixed point in the limit $\chi\rightarrow\infty$, we expect 
 $\xi_{\rm E}^{(n)}\rightarrow\xi^{\rm (n)}_{\rm MPS}$ in that limit. Down to $\chi=D^2$, 
 the two sets of (leading) correlation lengths remain quite similar so, in the following, we shall focus on $\xi_{\rm MPS}^{(n)}$ for clarity. 
 
\begin{figure}[htbp]
\centering
\includegraphics[width=0.99\columnwidth,angle=0]{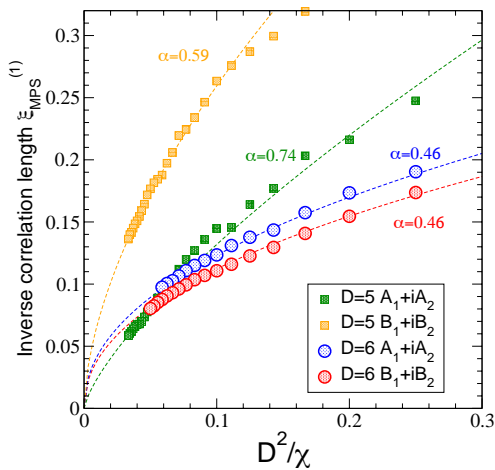}
	\caption{
Inverse of iPEPS (maximal) correlation length $1/\xi_{\rm MPS}^{(1)}$ (see text) versus $D^2/\chi$. The data for the four best  
optimized PEPS of bond dimension $D=5$ and $D=6$ are shown with different symbols according to legend.  Dashed lines are simple power-law fits, $1/\xi\simeq (D^2/\chi)^\alpha$, $\alpha<1$.
}
\label{fig:InvCorrLength}
\end{figure}

The maximal correlation length $\xi_{\rm MPS}^{(1)}$ has been computed 
in the best $D=5$ and $D=6$ 
PEPS (optimized up to $\chi=3D^2$) for increasing CTMRG dimension $\chi=kD^2$, $k\in \mathbb{N}^+$, up to 
$\chi=30D^2$ and $\chi=17D^2$, respectively. Results for $1/\xi_{\rm MPS}^{(1)}$ versus $D^2/\chi$ are shown in Fig.~\ref{fig:InvCorrLength}.  Power-laws fit the data well,
suggesting that the maximal correlation length diverges 
as $\xi_{\rm MPS}^{(1)}\propto \chi^\alpha$ when $\chi\rightarrow\infty$, with an exponent
$\alpha<1$ (slow divergence). Hence, surprisingly, the bulk seems critical, unlike the FQHS
analog. This is reminiscent of the spin-1/2 PEPS chiral spin liquid which also seems to be critical
(see comparison in Appendix~\ref{app0}). 

\begin{figure}[htbp]
\centering
\includegraphics[width=0.99\columnwidth,angle=0]{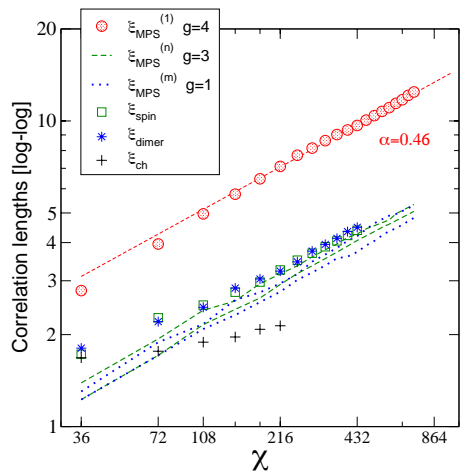}
	\caption{
Maximal and next four subleading correlation lengths $\xi_{\rm MPS}^{(n)}$, $n=1, ..., 5$, in the $D=6$ $B_1+iB_2$ PEPS (optimized up to $\chi=108$), versus $\chi$ in log-log scale, 
and sorted according to the degeneracy $g$ of the corresponding
TM eigenvalue $|t_n|$. The dashed (straight) line corresponds to a simple power-law divergence, 
$\xi=A \chi^\alpha$, $\alpha=0.46$. For comparison, spin and dimer correlation lengths are also shown.
}
\label{fig:CorrLength_loglog}
\end{figure}

To get more insights on the nature of the correlations in the PEPS chiral SL we 
have investigated the subleading correlation lengths $\xi_{\rm MPS}^{(n)}$, $n>1$. Since the $A_1+iA_2$ and $B_1+iB_2$ 
$D=6$ chiral PEPS have very similar properties, we shall focus, from now on, on the $B_1+iB_2$ $D=6$ PEPS.
Results for the largest five correlation lengths are plotted in Fig.~\ref{fig:CorrLength_loglog} on a log-log scale, showing a rather linear behavior over almost a decade. This confirms the (slow) power-law increase $\xi\propto\chi^\alpha$, $\alpha\simeq 0.46$, also for the subleading correlation lengths.

\subsubsection{From the real-space correlation functions}

\begin{figure}[htbp]
\centering
\includegraphics[width=0.99\columnwidth,angle=0]{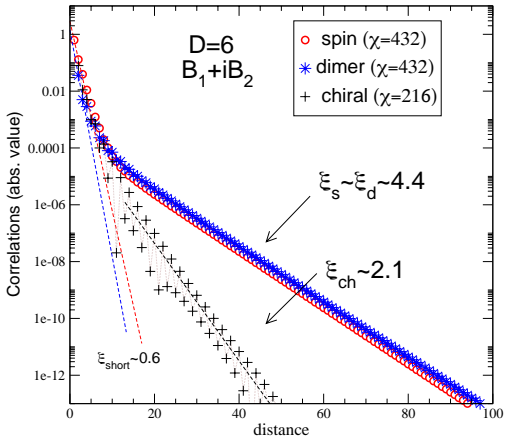}
	\caption{Spin-spin, dimer-dimer and chiral-chiral correlations (in absolute value) versus distance in the $D=6$, $B_1+iB_2$, chiral PEPS.  
Dashed lines are simple exponential fits of the short range decay of the spin-spin and dimer-dimer correlations.
The correlation lengths are extracted from the exponential decay at large distances.
}
\label{fig:CorrVsDist}
\end{figure}

In order to identify the type of physical operators these correlation lengths may be associated to, we have computed the spin-spin, (longitudinal) dimer-dimer and chiral-chiral correlations versus distance (see Appendix \ref{app1}
for details) and extracted the corresponding correlation lengths $\xi_{\rm s}$, $\xi_{\rm d}$ and $\xi_{\rm ch}$
from the long-distance behavior as illustrated in Fig.~\ref{fig:CorrVsDist}(a). 
We find that $\xi_{\rm s}$ and $\xi_{\rm d}$ are very close to the largest $\xi_{\rm MPS}$ with degeneracy
$g=3$ and $g=1$, respectively, consistent with triplet spin and singlet dimer operators. In contrast to $\xi_{\rm s}$ and
$\xi_{\rm d}$, the chiral correlation length grows very slowly suggesting that the chiral correlation remains short-range. Interestingly, the maximal correlation length
$\xi_{\rm MPS}^{(1)}$ is of degeneracy $g=4$ (which would naively correspond to a spin-3/2 operator) 
and hence cannot trivially be associated to a simple local operator acting on a group of physical sites but, perhaps, to chiral modes (see subsection~\ref{subsec:es_ring}) counter-propagating along the two chains of $T$ tensors of the long $(\mathcal{T}_{\rm MPS})^{\otimes N_h}$ ladder, $N_h\gg 1$. 

\begin{figure}[htbp]
\centering
\includegraphics[width=0.99\columnwidth,angle=0]{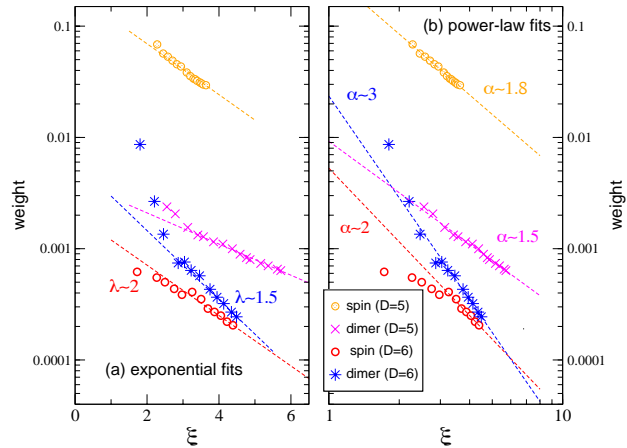}
	\caption{
	Weight associated to the exponential decay with the maximal correlation length, as a function of the
(maximal) correlation length, in the $B_1+iB_2$ chiral PEPS of bond dimension $D=5$ and $D=6$. 
(a) Semi-log plot. The straight lines are exponential fits $w(\xi)\sim \exp{(-\xi/\lambda)}$. (b) Log-log plot. The straight lines are power-law fits $w(\xi)\sim 1/\xi^\alpha$.
}
\label{fig:Weight_vs_chi}
\end{figure}

Let us now examine in more details the form and the magnitude of the spin-spin and dimer-dimer correlation functions 
at all length scales. 
First, at short distance, we observe a rapid exponential fall-off characteristic of the lattice Moore-Read spin liquid
(as for the spin-1/2 chiral PEPS, an ansatz for the lattice KL state). The length scale associated to this short range behavior
turned out to be very short, around $\xi_{\rm short}\sim 0.6$, as seen in Fig.~\ref{fig:CorrVsDist}.
More generally, one expects a sum of exponential contributions 
with a distribution of length scales. In other words, the  spin-spin (or dimer-dimer) 
correlation function vs distance can be written as a discrete sum,
\begin{equation}
C(d)=\sum_{\xi_{\rm short}\le\xi\le\xi_{\rm s}} w(\xi)\exp{(-d/\xi)}\, ,
\label{eq:Cs}
\end{equation}
where the short distance decay is characterized by $w(\xi_{\rm short})\simeq 1$ while, at long distance, 
the slower decay $\exp{(-d/\xi_{\rm s})}$
takes over. Typically, we find that $\xi_{\rm s}\gg \xi_{\rm short}$ and $w(\xi_{\rm s})\ll 1$. 
In the limit $\chi\rightarrow\infty$, one expects that the spectrum of the transfer matrix becomes dense, so that one can 
use a continuous integral over all eigenvalues for computing $C(d)$, namely $C(d)=\int_\xi d\xi n(\xi) w(\xi)\exp{(-\frac{d}{\xi})}$,
where $n(\xi)$ is the density of state. Fig.~\ref{fig:CorrLength_loglog} suggests that the density of eigenvalues is constant in
log-scale so that $n(\xi) d\xi \sim \frac{d\xi}{\xi}$.
In order to extract the possible functional form of the correlation function,  
it is now necessary to
get the behavior of the weight function $w(\xi)$.
To do so we have plotted 
$w(\xi_{\rm s})$ vs  $\xi_{\rm s}$ and $w(\xi_{\rm d})$ vs  $\xi_{\rm d}$ 
in Fig.~\ref{fig:Weight_vs_chi}(a,b), using semi-log and log-log scales. 
In fact, because of the limited range of available maximal correlation lengths (obtained by varying the environment dimension $\chi$), both exponential and power-law fits give reasonable results, hence providing different answers for the long-range correlations.
Let us examine each case separately.

{\it Exponential decay of $w(\xi)$ --} Let us assume $w(\xi)\simeq W_D\exp(-\xi/\lambda_D)$ in Eq.~(\ref{eq:Cs}) as a legitimate ansatz, where
$\lambda_D$ is a new emerging length scale and $W_D$ is some amplitude, which both depend on the PEPS bond dimension $D$. Typically, our fits in Fig.~\ref{fig:Weight_vs_chi}(a) give $\lambda_6\sim 2$ ($\lambda_6\sim 1.5$) for the spin-spin (dimer-dimer)
correlations of the $D=6$ chiral PEPS. If this functional form is correct then
$C(d)\sim W_D\int_\xi \frac{d\xi}{\xi} \exp{(-\frac{d}{\xi}-\frac{\xi}{\lambda_D})}$ will show a typical stretched exponential form at long distance,
\begin{equation}
C(d)\sim W_D \exp{\{-2(d/\lambda_D)^\frac{1}{2}\}} \, ,
\label{eq:tail1}
\end{equation}
up to a possible power-law prefactor in $d/\lambda_D$. Hence, in this case, the spin-spin and dimer-dimer
correlations  would decay faster than any power-law. 
Interestingly, the same functional form was also proposed for the spin-1/2 chiral PEPS~\cite{Poilblanc2017b}.

{\it Power-law decay of $w(\xi)$ --} Let us now assume $w(\xi)\simeq W_D/\xi^\alpha$ and substitute it in Eq.~(\ref{eq:Cs}).
In that case, a simple estimate of $C(d)\sim W_D\int_\xi \frac{d\xi}{\xi} \exp{(-\frac{d}{\xi})}\frac{1}{\xi^\alpha}$
gives a power-lay decay at long-distance of the form
\begin{equation}
C(d)\sim W_D \, (1/d)^\alpha \, .
\label{eq:tail2}
\end{equation}
From the fits in Fig.~\ref{fig:Weight_vs_chi}(b) one gets estimates of the exponents of the spin-spin and dimer-dimer
correlation functions, $\alpha_s\sim 2$ and $\alpha_d\sim 3$ respectively, for $D=6$.  

Note that, in deriving (\ref{eq:tail1}) and (\ref{eq:tail2}), we have omitted potential oscillatory $(-1)^d$ behavior of some of the exponential terms of
(\ref{eq:Cs}), which may reduce the range of the correlations.
In any case, it is clear that the  long-range tail of the spin-spin and dimer-dimer correlation functions, for a fixed bond dimension $D$,
is of quite small magnitude proportional to $W_D$. 
Moreover, from Fig.~\ref{fig:Weight_vs_chi}(a,b),
it is also clear that the magnitude $W_D$ of $w(\xi)$ decreases strongly by increasing $D$.
For example, $W_D$ associated to the spin correlation length gets smaller by two orders of magnitude, just by increasing
$D$ from 5 to 6. Hence, we conjecture that this ``gossamer'' long-range tail is a finite-$D$ artifact of the chiral PEPS which
should gradually disappear when increasing the bond dimension $D$.

\subsection{Spin structure factor}

The previous calculations of correlation lengths suggest the $D=6$ chiral PEPS exhibits a form of long range tail, i.e. with a slower decay than a pure exponential function.
In the case of a power-law decay of the spin-spin correlations like 
$C_{\rm s}({\bf r})=W_s (1/|{\bf r}|)^\alpha \exp{(i{\bf Q}_{\rm AF}\cdot{\bf r})}$ (the antiferromagnetic wavevector 
${\bf Q}_{\rm AF}=(\pi,\pi)$ is consistent with the data), the static spin structure factor $S({\bf q}) =\sum_i \exp{(i{\bf q}\cdot{\bf r_i})} C_s ({\bf r_i})$
would diverge when ${\bf q}\rightarrow{\bf Q}_{\rm AF}$, only if $\alpha<2$. 
In contrast, in the case of a stretched exponential form, as suggested by the data, no divergence is expected at any momentum.

\begin{figure}[htbp]
\centering
\includegraphics[width=0.99\columnwidth,angle=0]{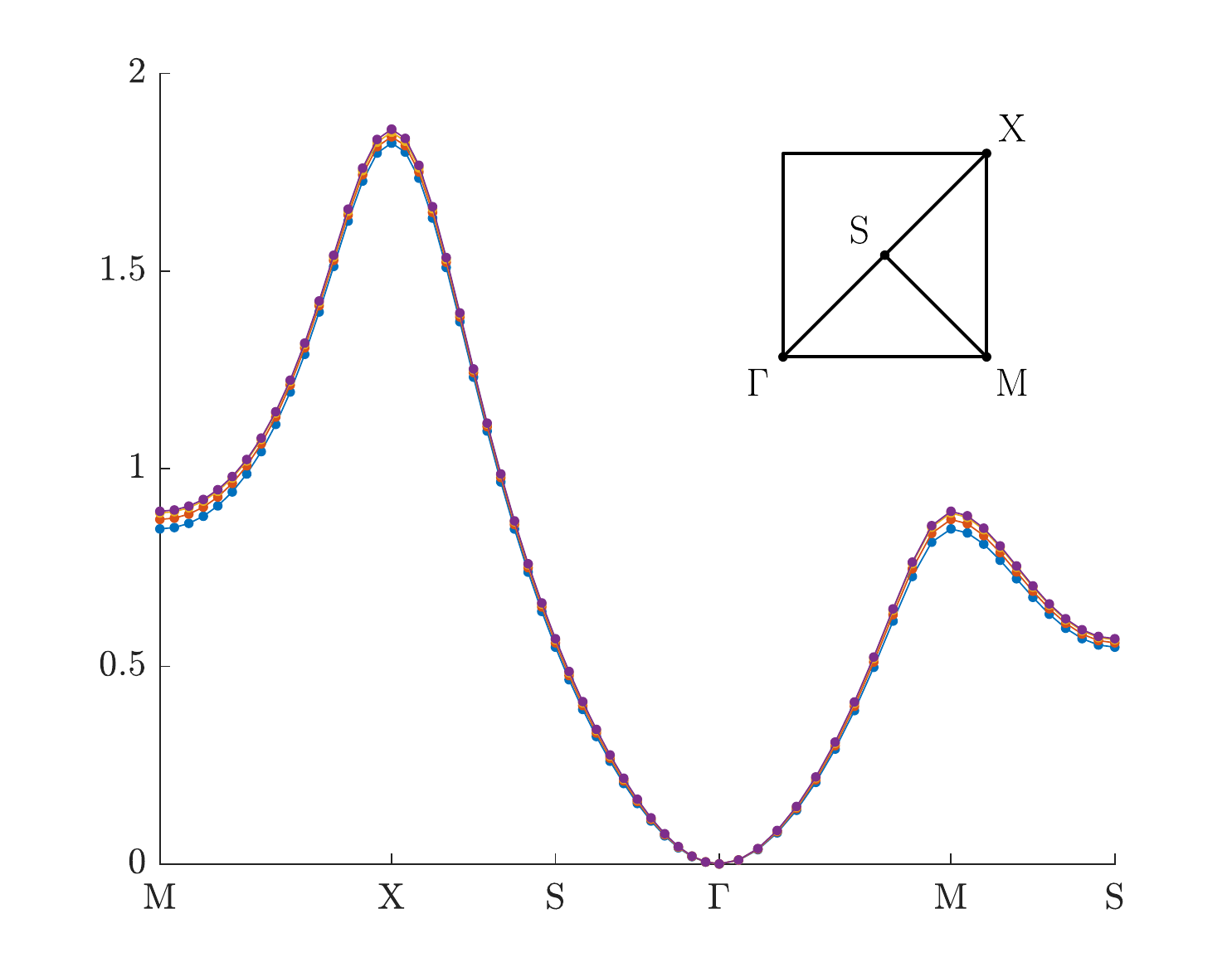}
	\caption{
Structure factor of the $D=6$ $B_1+iB_2$ chiral PEPS along the path in the Brillouin zone shown in the inset, for different values of the channel environment bond dimension $\chi=87$ (blue), $\chi=131$ (red), $\chi=158$ (yellow), and $\chi=185$ (purple). Due to finite-$\chi$ effects we find slightly negative values (of the order of the truncation error) at the $\Gamma$ point; in order to better compare the different $\chi$ we have shifted the curves by this small offset.}
\label{fig:SofQ}
\end{figure}

In order to get more insight, we have computed the spin structure factor in the $D=6$ chiral PEPS, and results are 
shown in Fig.~\ref{fig:SofQ}. The results do not show any sign of a divergence at any momentum,
suggesting that the decay of the spin-spin correlations at long distance 
is relatively rapid and corresponds to a ``gossamer'' tail,
consistent either with (\ref{eq:tail1}) or with (\ref{eq:tail2}), provided $\alpha \ge 2$.

\section{Edge modes}
\label{sec:edge}

\subsection{Expected SU(2)$_2$ counting}

As suggested by Li and Haldane~\cite{Li2008}, the entanglement spectrum (ES) offers a very powerful tool
to identify topological order in Abelian and non-Abelian liquid states since the latter is in one-to-one correspondence 
with the edge states. Since our PEPS is expected to provide a reliable lattice representation of the Moore-Read 
FQH state (apart from -- we believe -- a spurious long range ``gossamer" tail in some bulk correlations), the edge modes should be described by a SU(2)$_2$ WZW theory for which the ES will provide indisputable 
finger prints. The SU(2)$_2$ WZW theory harbors 3 sectors whose towers of states -- obtained by combining a bosonic 
(harmonic oscillator) mode with an independent Majorana-fermion mode -- are listed in Table~\ref{Table:su2_2}.

\begin{table}[htb]
\begin{center}
\resizebox{0.95\columnwidth}{!}{%
  \begin{tabular}{@{} cccc @{}}
    \hline \hline
    $n \backslash j $ & 0 & $\frac{1}{2}$ & 1 \\ 
  \hline 
  & & &\\
0 & (0) & ($\frac{1}{2}$)  &(1) 
\\
1 & (1)& ($\frac{1}{2}$)+($\frac{3}{2}$)& (0)+(1) 
\\
2 & (0)+(1)+(2)  & 2($\frac{1}{2}$)+2($\frac{3}{2}$)  & (0)+2(1)+(2)
\\
3 & (0)+3(1)+(2)  & 4($\frac{1}{2}$)+3($\frac{3}{2}$)+($\frac{5}{2}$) & 2(0)+3(1)+2(2)
\\
4 & 3(0)+4(1)+3(2)  & 6($\frac{1}{2}$)+6($\frac{3}{2}$)+2($\frac{5}{2}$) & -
\\
5 & 3(0)+8(1)+4(2)+(3)  & 10($\frac{1}{2}$)+10($\frac{3}{2}$)+4($\frac{5}{2}$) & -
\\
& & &\\
    \hline 
    \hline
      \end{tabular}
     }
\caption{[Color online] Towers of states of the SU(2)$_2$ WZW model, in each of the three sectors characterized by
the primary fields $j=0,\frac{1}{2},1$ (listed in each column) and conformal weights $\frac{1}{4}j(j+1)$. Each line corresponds to a Virasoro 
level indexed by $n$. For each sector and each level, the (quasi-) degenerate states can be grouped in terms of exact SU(2) multiplets
like $n_0 (0) + n_1 (1) +\cdots$ (meaning $n_0$ singlets, $n_1$ triplets, etc...). }
\label{Table:su2_2}
\end{center}
\end{table}

\subsection{Bulk-edge correspondence}

Let us now consider the chiral PEPS $|\Psi_{\rm PEPS}\big>$ on an infinitely long (horizontal) cylinder of finite circumference $N_v$, and a bipartition along a vertical cut into two right (R) and left (L) semi-infinite cylinders. One can then define the reduced density matrix (RDM) 
$\rho$ of, let say, the L part by taking the trace of $|\Psi_{\rm PEPS}\big>\big<\Psi_{\rm PEPS}|$ over the (physical) degrees of freedom of the R part.
Rewriting the positive operator $\rho$ as $\exp{(-H_b)}$, where $H_b=-\ln{\rho}$ can be viewed as 
a ``boundary'' Hamiltonian, Li and Haldane conjectured~\cite{Li2008} that the spectrum of $H_b$ -- dubbed
entanglement spectrum -- is in one-to-one correspondence with the actual edge spectrum of the partitioned system. Therefore, the ES should exhibit crucial information on the nature of the chiral edge modes which, in turn, can provide a precise characterization of our chiral SL. 

\begin{figure}[htb]
\centering
	\centering
\hfill\subfloat[]{\includegraphics[width=0.23\columnwidth,angle=0]{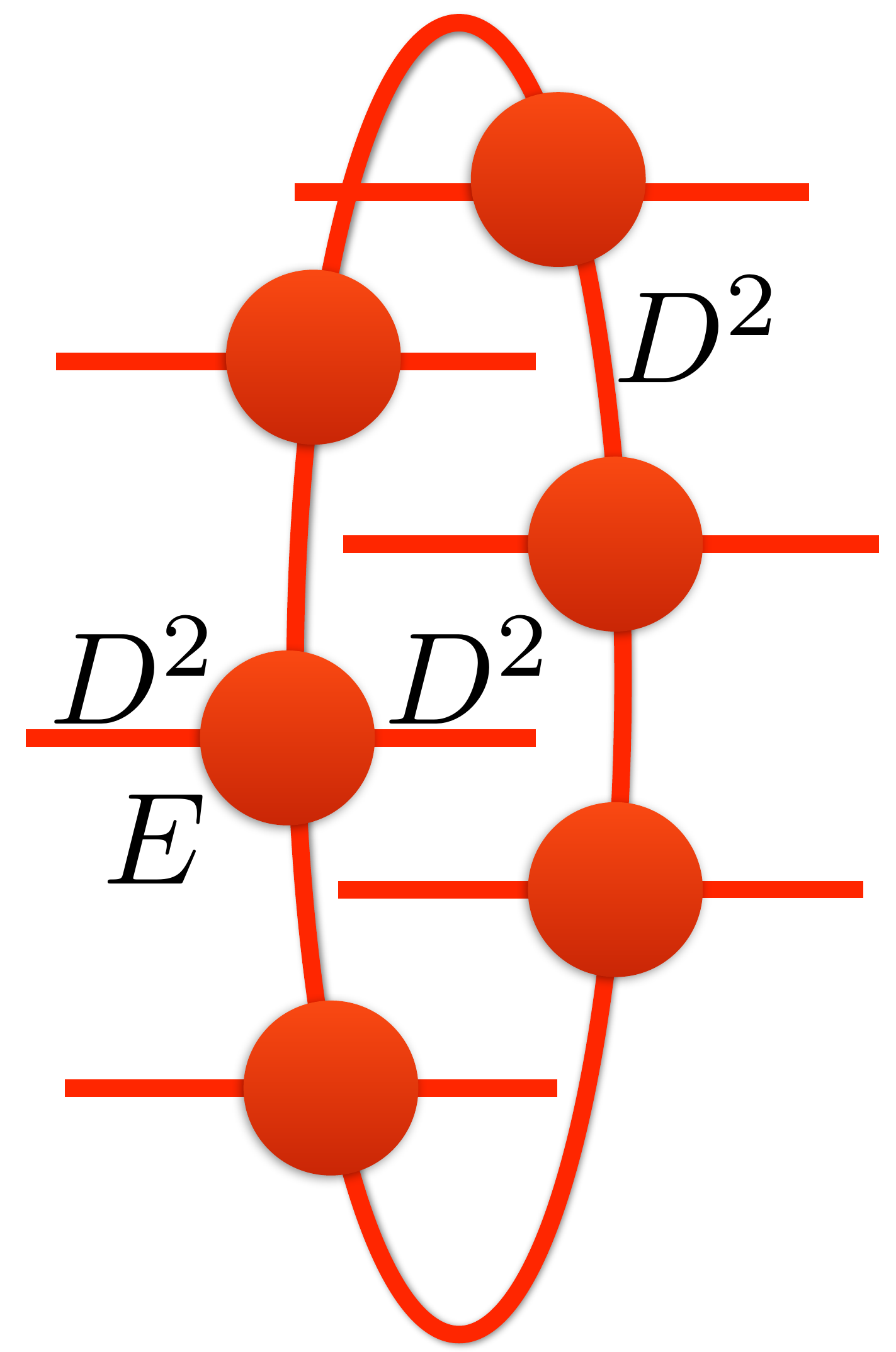}}
\hfill\subfloat[]{\includegraphics[width=0.5\columnwidth,angle=0]{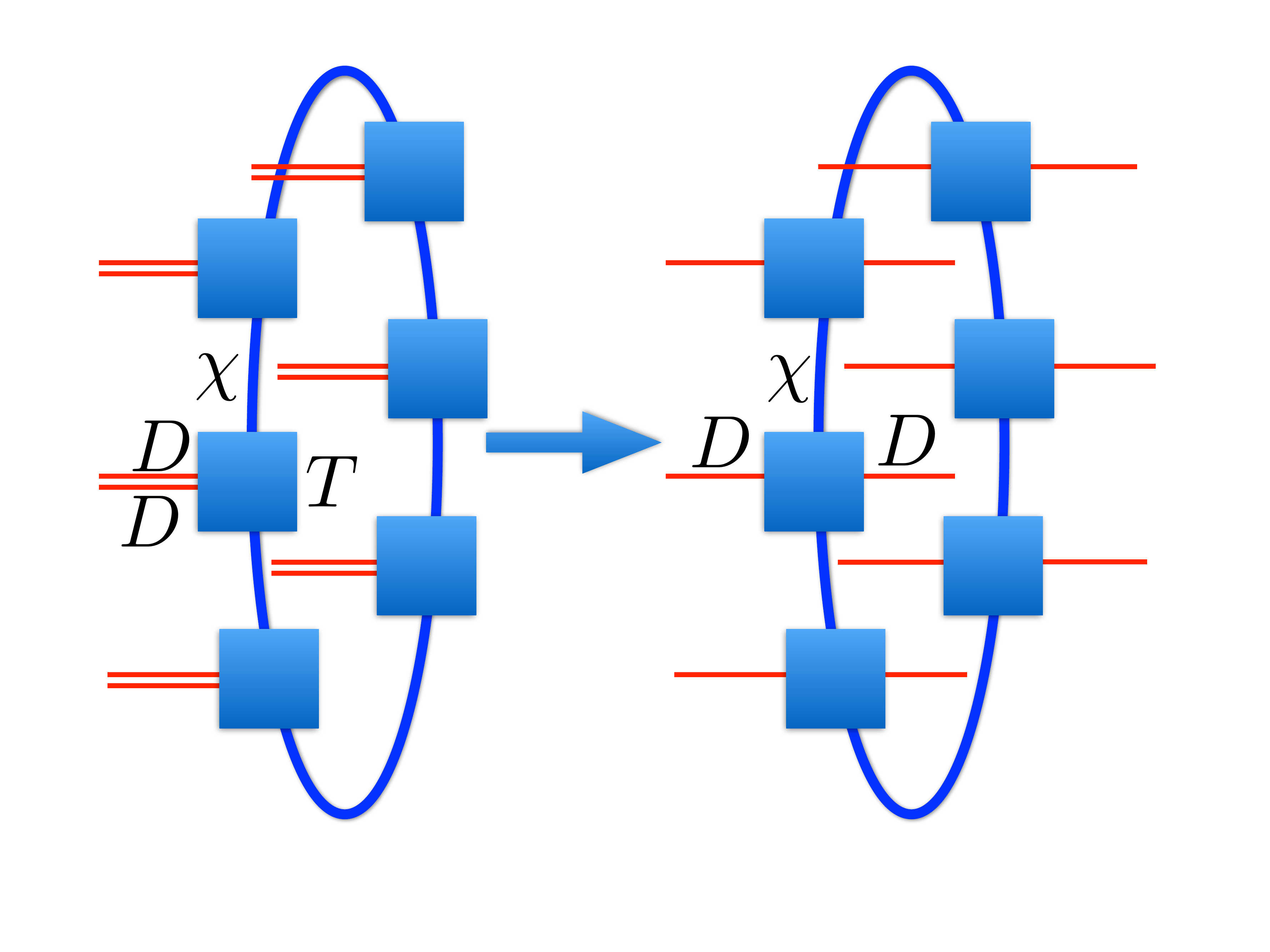}}\hfill
\caption{
 (a) Cylinder transfer matrix. (b) Boundary (right) vector obtained by contracting $N_v$ ($=6$ here) environment $T$ tensors on a ring is reshaped into the RDM $\rho_R$.}
\label{fig:RDM}
\end{figure}

To compute the ES on a cylinder, one can use the PEPS bulk-edge correspondence theorem~\cite{Cirac2011} that provides an exact relation between the RDM -- whose support is the 2D half-cylinder {\it physical} Hilbert space -- and a 1D (positive) operator $\sigma_b^2$ only acting on the $D^{N_v}$ {\it virtual} degrees of freedom of the "cut". The above fundamental relation involves an isometry that preserves the spectrum 
and~\cite{Cirac2011}
\begin{equation}
\sigma_b^2=\sqrt{\sigma_L^T}\sigma_R\sqrt{\sigma_L^T}\, ,
\label{eq:bulk-edge}
\end{equation}
where $\sigma_L$ ($\sigma_R$) is obtained from the cylinder TM (shown in Fig.~\ref{fig:RDM}(a)) left (right) leading eigenvector of dimension $(D^2)^{N_v}$, reshaped as a $D^{N_v}\times D^{N_v}$ matrix. 
Here, $\sigma_L^T=\sigma_R$ so that $\sigma_b=\sigma_R$.  
In previous work on spin-1/2 chiral spin liquids~\cite{Poilblanc2015}, $\sigma_b^2$ was obtained for 
a simple chiral $D=3$ PEPS 
by exact tensor contractions, multiplying recursively the cylinder TM to get the leading eigenvector.
For our $D=6$ chiral PEPS such a procedure is no longer possible and one has to rely on approximate contraction schemes. Below we describe two methods to compute $\sigma_b$, either using our previous knowledge of the
iPEPS environment or using a uniform MPS implementation. 

\subsection{Finite $N_v$ calculation using CTMRG  }
\label{subsec:es_ring}

First, we construct the half-cylinder leading eigenvector using the previously obtained $T$ tensor, (i) by contracting $N_v$ such tensors on a ring of $N_v$ sites, and (ii) by reshaping this vector into a $\sigma_b$ matrix, as shown in Fig.~\ref{fig:RDM}(b).
In this procedure, the CTMRG   parameter $\chi$ becomes the only control parameter of the approximate calculation of the ES for finite $N_v$. Note however that, strictly speaking, the calculation does not become exact in the 
$\chi\rightarrow\infty$ limit since $T$ is computed for an infinite system. However, this procedure may be more advantageous in the sense that it may reduce some of the finite-size effects in the ES, in comparison to the exact contraction method. 

In order to compute the ES, we first block-diagonalize the edge operator $\sigma_b$ using the exact symmetries of the PEPS, namely
(i) its $\mathbb{Z}_2$ gauge symmetry -- leading to even and odd sectors -- (ii) its full $SU(2)$-spin rotational invariance -- leading to 
sectors labelled by the total spin projection $S_z$ and (iii) its translational invariance -- leading to sub-sectors labelled by the
edge momentum $K=n\frac{2\pi}{N_v}$, $n\in\mathbb{Z}$. 
First, as for the spin-1/2 chiral PEPS~\cite{Poilblanc2015,Poilblanc2016}, the even (odd) topological gauge sector only contains the integer (half-integer) $S_z$ sectors. Secondly, because of the chosen $\pi$-spin rotation on the sites of the B-sublattice (in order to deal with the same unique tensor on the A and B sites), the $S_z$ operator acquires a minus sign under an odd number of lattice translations, so that only (i) $|S_z|$ and $K$ or (ii) $S_z$ and $K$ [mod $\pi$] can be considered simultaneously. 
 
\begin{figure}[htbp]
\centering
\includegraphics[width=0.99\columnwidth,angle=0]{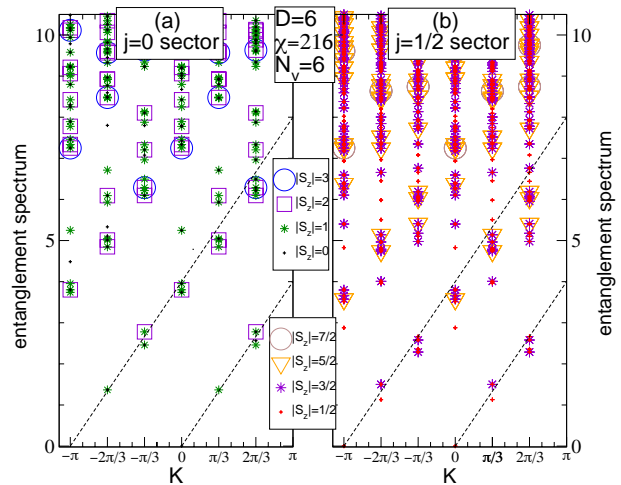}
	\caption{
Entanglement spectrum of the $B_1+iB_2$ $D=6$ chiral PEPS for $N_v=6$ (see text), versus edge momentum $K$, computed 
for $\chi=216$.
Different symbols correspond to different
values of  $|S_z|$, showing that the spectrum is composed of exact $SU(2)$ multiplets with integer (a) and half-integer spins (b). 
Dashed lines correspond to the low-energy chiral CFT modes. }
\label{fig:ES_chi144_Sz}
\end{figure}

\begin{figure}[htbp]
\centering
\includegraphics[width=0.99\columnwidth,angle=0]{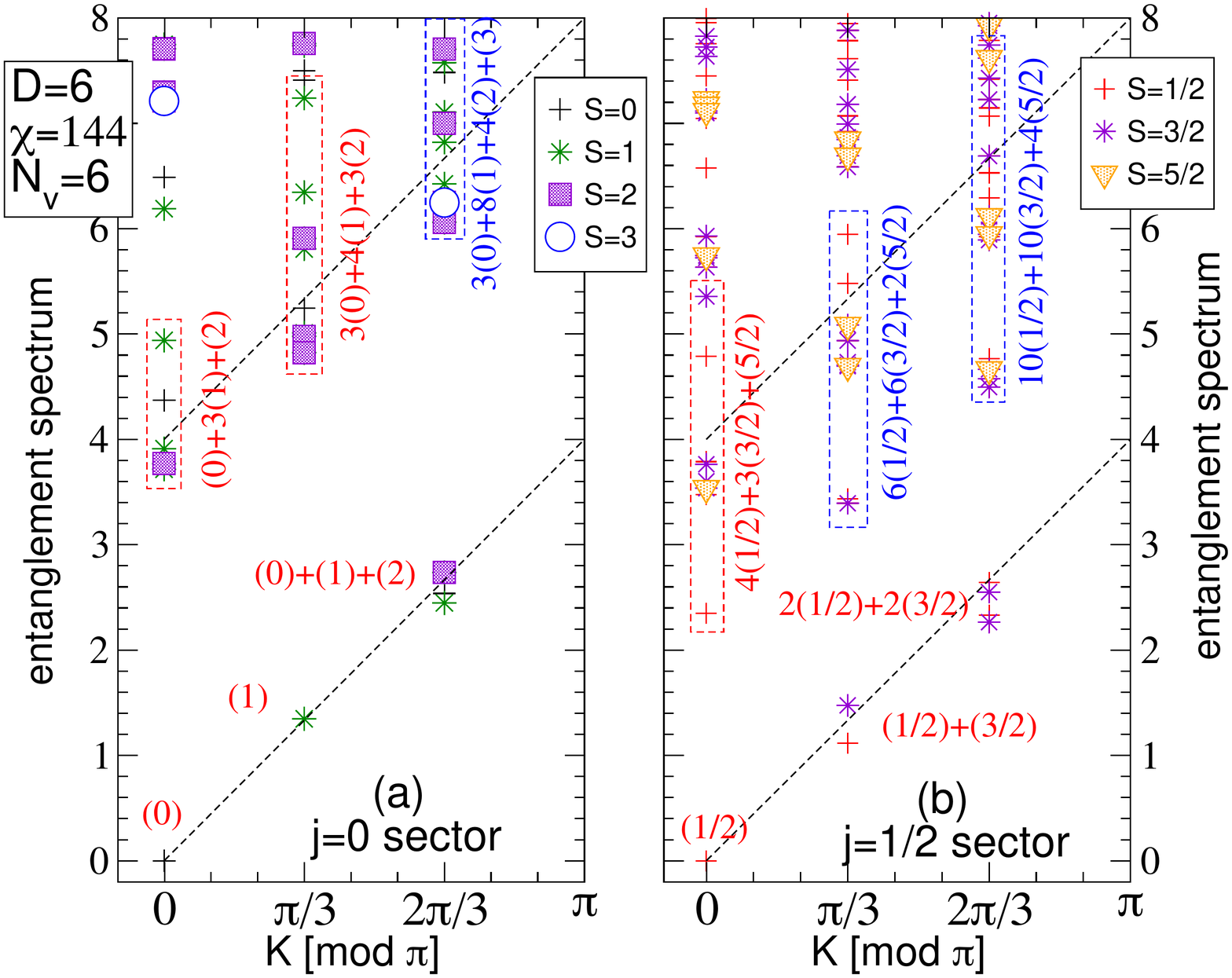}
\includegraphics[width=0.99\columnwidth,angle=0]{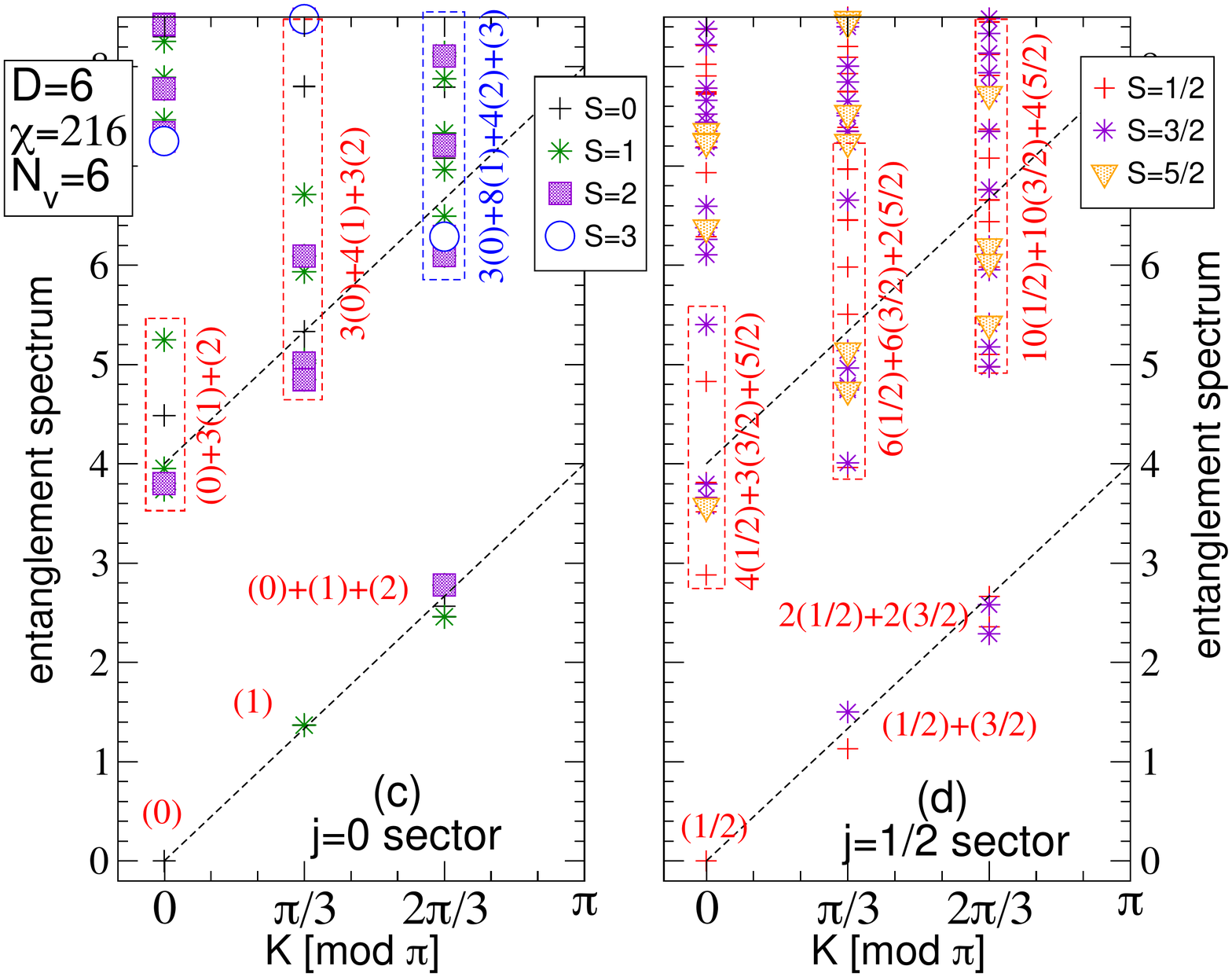}
	\caption{
Close-up of the low (quasi-)energy entanglement spectrum of the $D=6$ chiral PEPS for $N_v=6$ versus $K$ [mod $\pi$]
(see text). Comparison of the spectra for $\chi=144$ (a,b) and $\chi=216$ (c,d) is shown showing an almost convergence with $\chi$. 
The expected $j=0$ and $j=1/2$ chiral modes of the 
WZW $SU(2)_2$ theory appear at the bottom of the $\mathbb{Z}_2$-even (a,c) and $\mathbb{Z}_2$-odd (b,d)  sectors of the ES, respectively. The SU(2)-multiplet content of each group of levels is indicated, in agreement with the CFT prediction (except for the blue boxes where a few levels are missing).}
\label{fig:ES_chi144}
\end{figure}

The entanglement spectrum of the $D=6$ $B_1+iB_2$ PEPS has been computed on a $N_v=6$ site ring 
from the $T$ environment tensor with $\chi=D^2=36$,  $\chi=4D^2=144$ and $\chi=6D^2=216$.
Data for $\chi=216$ are shown in Fig.~\ref{fig:ES_chi144_Sz} in the range
$[-\pi,\pi)$ (the complete spectrum for $\chi=144$ is also shown in Fig.~\ref{fig:ES_chi144_ALL} in the Appendix~\ref{app2}).  Two low
(quasi-)energy chiral branches emerge, linearly dispersing
in one direction only, separated by momentum $\Delta K=\pi$. By grouping together the degenerate energy 
levels at $K$ and $K+\pi$, one obtains exact 
SU(2) multiplets that can be labeled by their total spin $S$ quantum 
number~\footnote{Note that the ES eigenvalues in the odd sector of the half-integer spin multiplets
are all exactly two-fold degenerate as for the spin-1/2 chiral spin liquid~\protect\cite{Poilblanc2015,Poilblanc2016} due to an interplay between SU(2) and space-group symmetries~\protect\cite{Hackenbroich2018}}. The same spectrum can then be 
plotted for $K\in [0,\pi[$, mod $\pi$, labelling now the levels according to their spin $S$,
as shown in Fig.~\ref{fig:ES_chi144}. In this way, in each topological sector, the two branches 
merge into a unique chiral branch composed of groups of quasi-degenerate exact SU(2) multiplets, 
labeled as $\oplus_{S_{\rm min}}^{S_{\rm max}} n_S (S)$, $n_S\in\mathbb{N}$. Examining carefully each group of multiplets, for increasing momentum $K=n\frac{2\pi}{N_v}$, we find that their content agrees exactly -- at least up to the fourth level -- the prediction of the SU(2)$_2$ WZW conformal field theory of central charge $c=3/2$, characterized by a bosonic mode combined with an Ising anyon (or ``Majorana fermions''). A comparison with the ES of the spin-1/2 CSL is shown in Appendix~\ref{app2}, showing a very distinct SU(2)$_1$ CFT counting. 

Note that the third $j=1$ topological tower cannot be derived straightforwardly since, it probably requires the insertion of a string of 
$\mathbb{Z}_2$ ``vison" operators in the cylinder direction (see Ref.~\onlinecite{Poilblanc2015} for the case of the simple
spin-1/2 CSL). The representation of the vison operator in the dimension-$\chi$ fixed-point basis is not known. 

\subsection{Uniform MPS calculation}

\begin{figure}[htb]
\centering
	\centering
\hfill\subfloat[]{\includegraphics[width=0.49\columnwidth,angle=0]{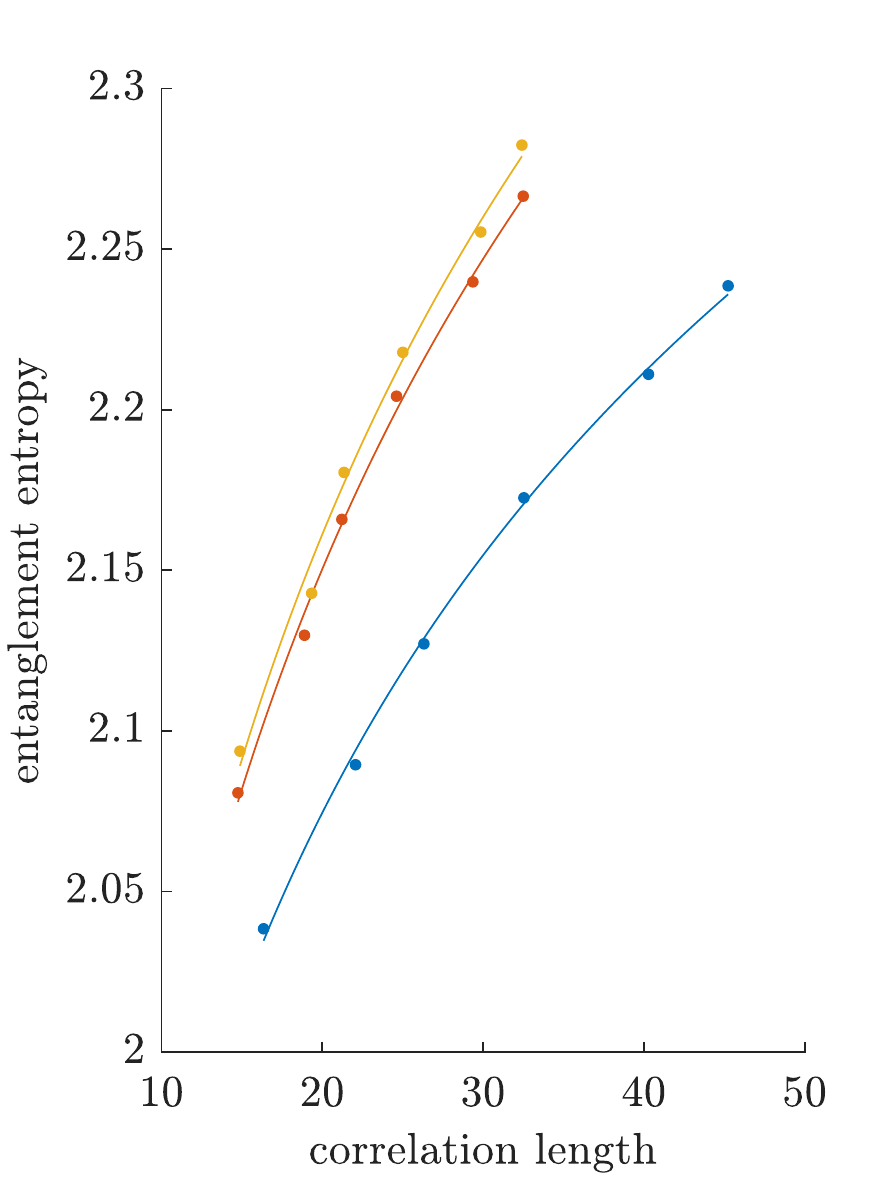}} 
\hfill\subfloat[]{\includegraphics[width=0.5\columnwidth,angle=0]{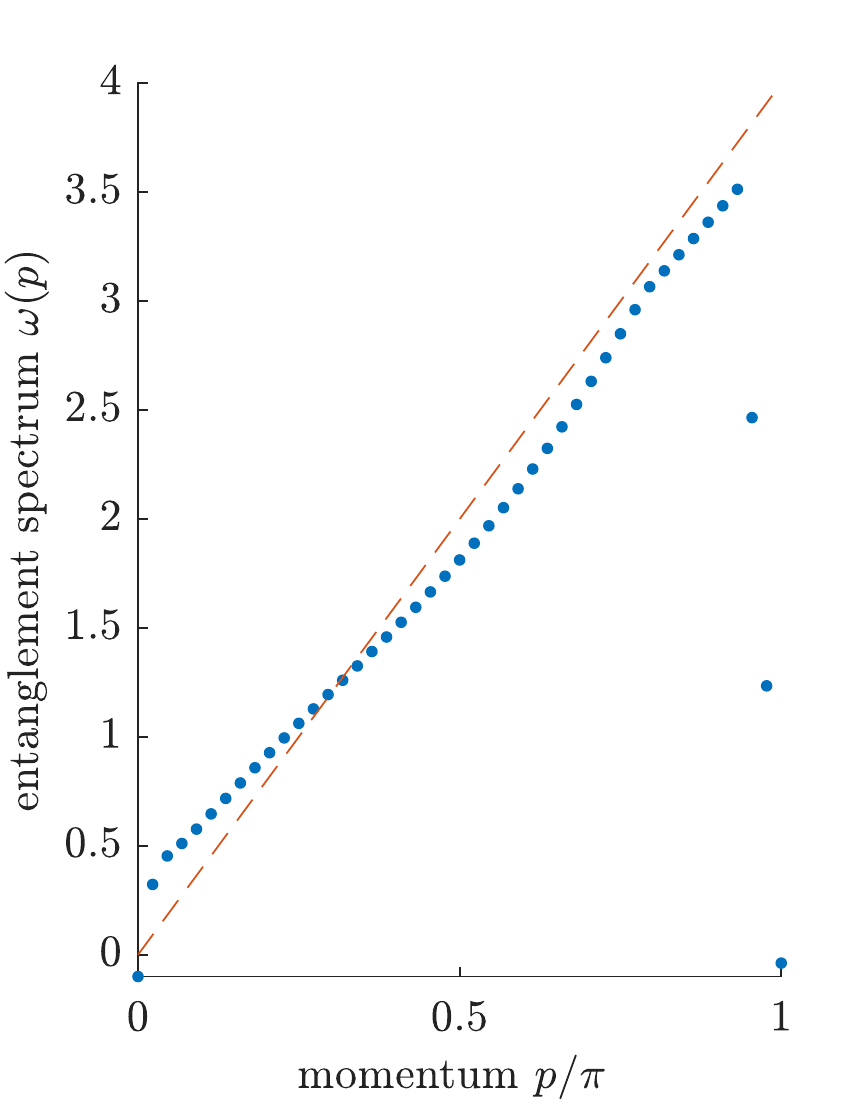}} \hfill
\caption{
 (a) The bipartite entanglement entropy as a function of the correlation length for MPS approximations (with increasing bond dimension $\chi'=50\to100$) of the fixed point of $\rho\approx\exp(-H_b)$; the three data sets are for different values of the bond dimension $\chi$ of the MPO approximation for $\rho$: $\chi=87$ (blue), $\chi=131$ (red) and $\chi=158$ (yellow). The lines are fits to the data points with $S\propto\frac{c}{6}\log(\xi)$, where we find $c_{\chi=87}=1.187$, $c_{\chi=131}=1.429$, $c_{\chi=158}=1.462$. (b) The spectrum of $H_b$ as computed with the quasiparticle ansatz on the MPO approximation $\rho\approx\exp(-H_b)$ with bond dimension $\chi=158$ in the thermodynamic limit.}
\label{fig:mps_boundary}
\end{figure}

We can also characterize the entanglement spectrum using uniform MPS techniques. We take the fixed point of the PEPS transfer matrix, interpret it as a matrix-product operator $\rho$ (with bond dimension $\chi$) representing the boundary Hamiltonian as $\rho=\exp(-H_b)$ in the thermodynamic limit directly. The fixed point of this MPO corresponds to the ground state of the boundary Hamiltonian $H_b$, which we can, again, approximate as a uniform matrix product state. In particular, we can plot the scaling of the bipartite entanglement entropy of this uniform MPS as a function of the its correlation length, which is known \cite{Pollmann2009} to be related to the central charge as $S\propto\frac{c}{6}\log(\xi)$. As shown in Fig.\ref{fig:mps_boundary}(a), this provides clear evidence that the boundary theory is described by a CFT with central charge $c=3/2$. In addition, we can compute the spectrum of $H_b$ by applying the quasiparticle excitation ansatz \cite{Haegeman2016}. In Fig.~\ref{fig:mps_boundary}(b) we have plotted the entanglement spectrum, showing the signature of a chiral spectrum. The presence of a very steep "left-moving" branch and a small modulation on top of the linear dispersion are the consequence of the finite-$\chi$ approximation of $\rho$.

\section{Summary, discussion and outlook} 

Our iPEPS study provides the first investigation of a (fine-tuned) spin-1 frustrated Heisenberg model  on the square lattice,
which includes a time-reversal breaking
plaquette term. The ES and the scaling of the entanglement entropy provide smoking gun evidence of SU(2)$_2$ chiral edge modes 
with central charge $c=3/2$, consistent with a bulk non-Abelian CSL realizing, on the lattice, the Moore-Read FQH state. 

However, our results also point towards long-range behavior of some bulk correlations (such as spin-spin or dimer-dimer correlation functions) which may be algebraic or more rapidly decaying (as e.g. for a stretched exponential). We argue that, although this behavior might be generic in a chiral PEPS (see the exact proof for
Gaussian PEPS in Ref.~\onlinecite{Dubail2015}) it is a spurious artifact which {\it does not constitute a real obstruction} for accurate investigation of truly gapped CSL ground states of simple (frustrated) quantum spin Hamiltonians. 
First, we note that chiral edge modes can be truly gapless only if the effective 
one-dimensional PEPS boundary Hamiltonian $H_b$  (acting on the virtual space) is long range, which in turn implies, from the PEPS bulk-edge correspondence~\cite{Cirac2011}, that the (maximal) bulk correlation length indeed diverges. However, our results also show that this divergence is connected to a long-range  tail in some physical correlation functions, whose magnitude is already quite small for $D=6$ and seems to decrease rapidly
with the tensor bond dimension. This suggests that this ``gossamer'' tail will gradually disappear when $D\rightarrow\infty$, even with no practical effect to faithfully approximate a gapped CSL (as far as energy, short and intermediate range correlations, edge mode physics, etc... are concerned) with a finite-$D$ chiral PEPS. Therefore, the PEPS formalism seems to be 
an unbiased efficient method to investigate other non-Abelian CSL in higher spin SU(2)-invariant HAFM (with or without explicit time-reversal breaking) and in SU(N) models. 

Lastly, we comment on possible experimental realizations of Hamiltonian (\ref{eq:model}).
Let us start from a two-orbital Hubbard model with on-site Hubbard and (ferromagnetic) Hund couplings, and hoppings on both NN and NNN sites, respectively. In some limit of very large 
Hubbard and Hund couplings, only localized spin-1 degrees of freedom can be retained on the sites and
two-site spin interactions appear in second order in the hoppings. 
Moreover, if an orbital flux is including in the plaquettes of the square lattice 
(breaking time-reversal symmetry), then chiral terms appear in third order of the hoppings. 
As suggested in Refs.~\onlinecite{Nielsen2013,Nielsen2014} (although for the spin-1/2 case), such Hamiltonians
can be realized using e.g. ultracold atoms loaded on optical lattices in the presence of a synthetic 
gauge field~\cite{Gross2017}.

\section*{acknowledgements}

This project is supported in part by the TNSTRONG
ANR grant (French Research Council).  This work was granted access to the HPC resources of CALMIP and GENCI supercomputing centers under the allocation 2017-P1231 and A0030500225, respectively. LV was supported by the Flemish Research Foundation.  We acknowledges inspiring conversations with F. Becca, A. L\"auchli, G. Misguich, P. Pujol, G. Sierra, S. Simon and A. Sterdyniak. We are also thankful to B.~Estienne for 
help in establishing the content of the SU(2)$_2$ CFT tower of states.

\bibliography{bibliography}

\appendix
\renewcommand\thefigure{\thesection.\arabic{figure}}
\setcounter{figure}{0}

\section{Comparison of the correlation lengths in the spin-1 and the spin-1/2 chiral HAFM}
\label{app0}

\begin{figure}[htbp]
\centering
\includegraphics[width=0.99\columnwidth,angle=0]{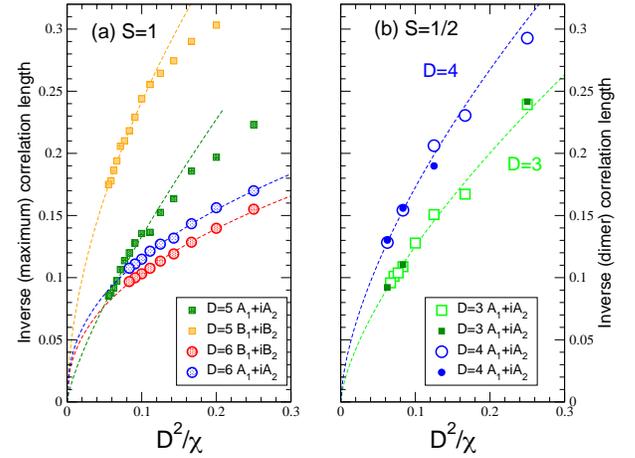}
	\caption{
Comparison between the inverse iPEPS (maximal) correlation length $\xi_{\rm E}^{(1)}$ (see text) of the spin-1 chiral HAFM (a) and
the inverse iPEPS (dimer) correlation length $\xi_{\rm dimer}$ of the spin-1/2 chiral HAFM (b). The data for the four best  
optimized spin-1 PEPS of bond dimension $D=5$ and $D=6$ are shown in (a). In (b), open and closed symbols correspond to two different sets of the Hamiltonian parameters given by $J_1=K_c=1$, $J_2$=0 (open symbols) and
$J_1=0.89$, $J_2=0.42$ and $K_c=0.375$ (closed symbols). 
Dashed lines are simple power-law fits, $1/\xi\simeq (D^2/\chi)^\alpha$, $\alpha<1$.}
\label{fig:CorrLength_comp}
\end{figure}

The maximal correlation length $\xi_{\rm E}^{(1)}$ of the spin-1 chiral PEPS has been obtained from the two largest eigenvalues of the 
$\mathcal{T}_{\rm E}$ transfer matrix depicted in Fig.\ref{fig:CTMRG}(g). Its divergence 
with $\chi$ was shown to track the one of the dimer-dimer correlation length. 
In Fig.~\ref{fig:CorrLength_comp} we compare $\xi_{\rm E}^{(1)}$ in the spin-1 $D=5$ and $D=6$ chiral PEPS
to the dimer-dimer correlation length in the related spin-1/2 chiral HAFM,
for two choices of the model parameter considered in Ref.~\cite{Poilblanc2017b}. The behaviors of the spin-1/2 and spin-1 chiral PEPS
are very similar,
both being consistent with a power-law divergence (as shown by the dashed line fits). 

\section{Computing correlation functions with iPEPS}
\label{app1}

\begin{figure}[htb]
\centering
	\centering
\hfill\subfloat[]{\includegraphics[width=0.42\columnwidth,angle=0]{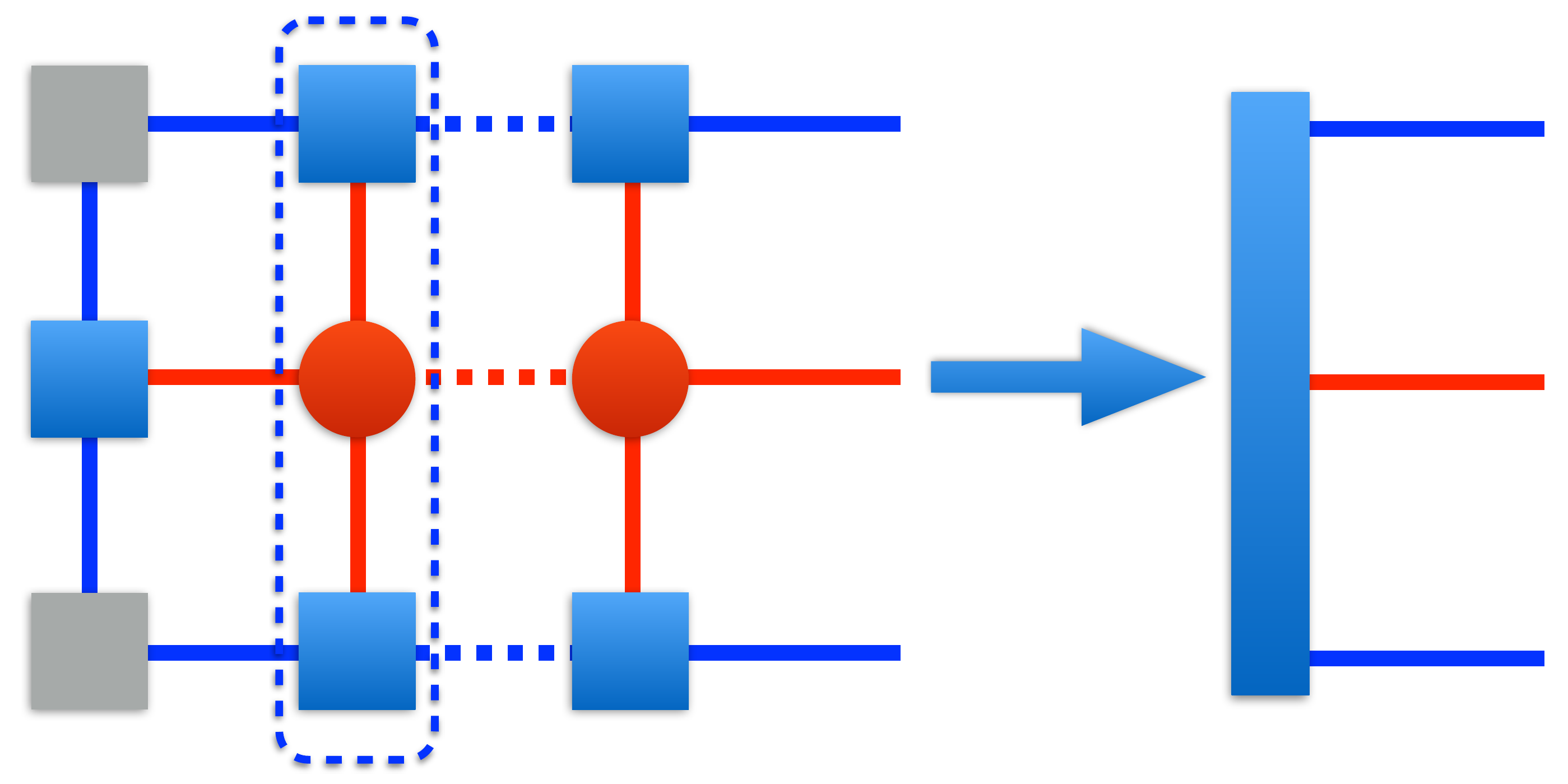}}
\hfill\subfloat[]{\includegraphics[width=0.42\columnwidth,angle=0]{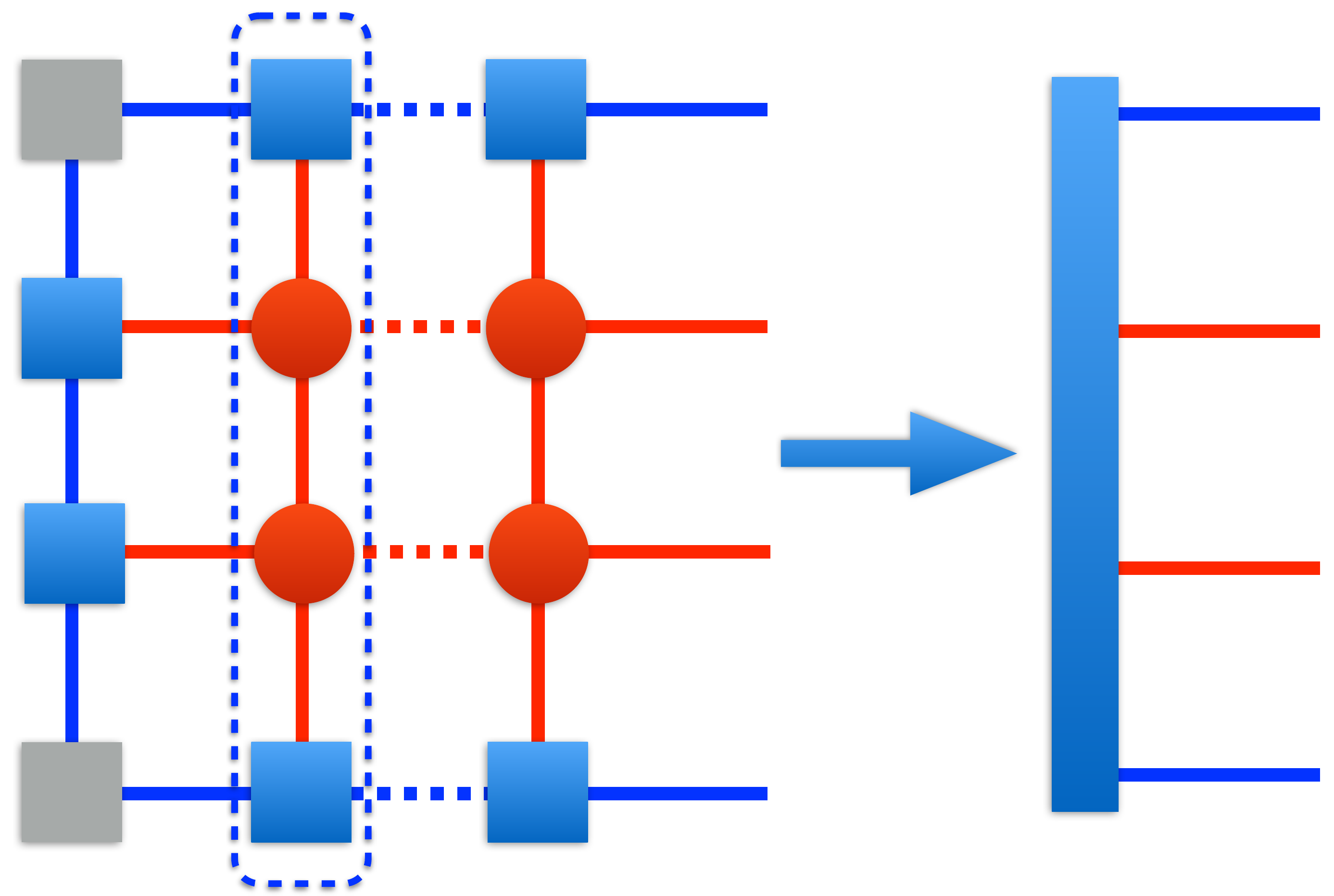}} \hfill\\

\hfill\subfloat[]{\includegraphics[width=0.38\columnwidth,angle=0]{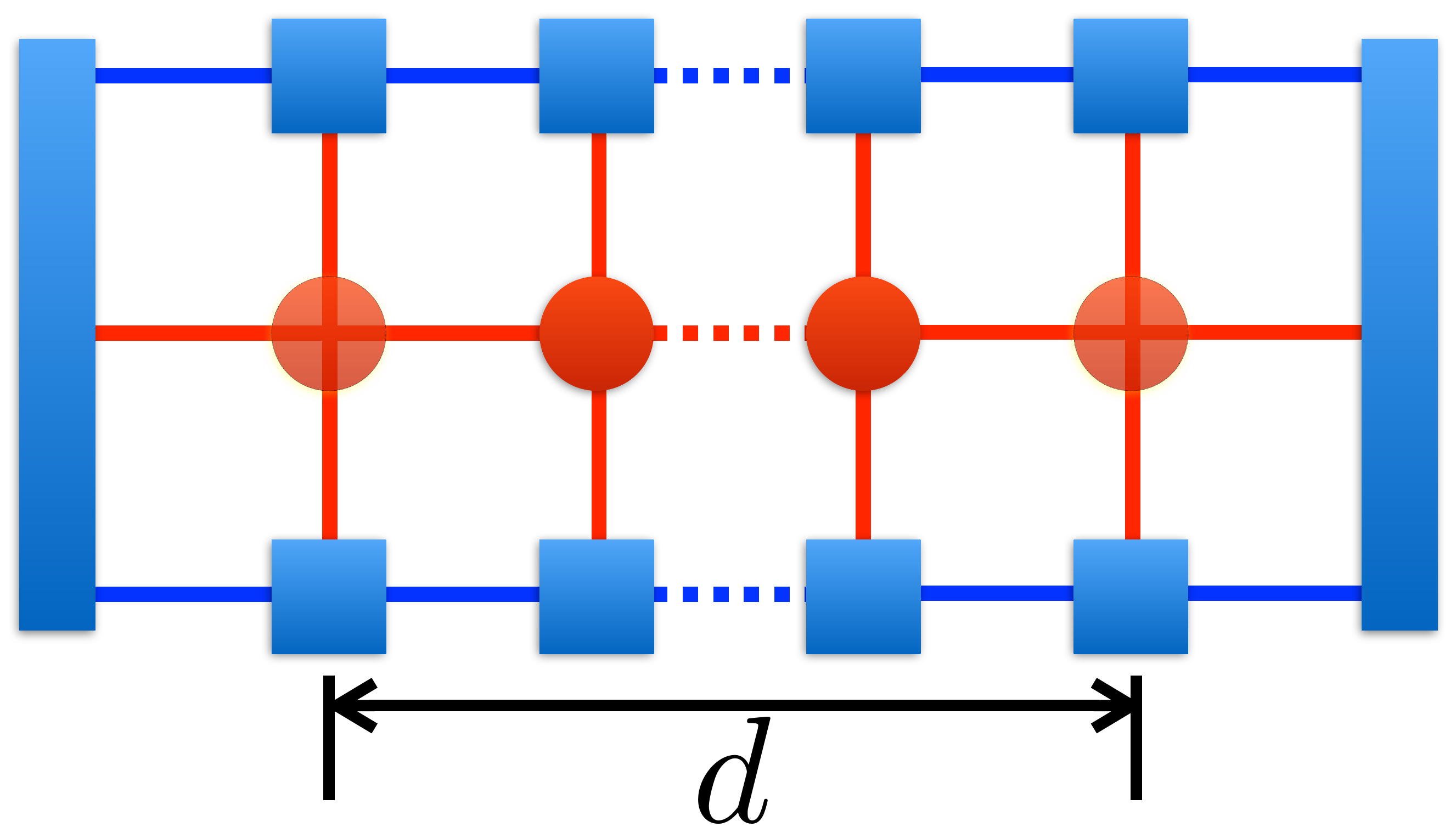}}
\hfill\subfloat[]{\includegraphics[width=0.47\columnwidth,angle=0]{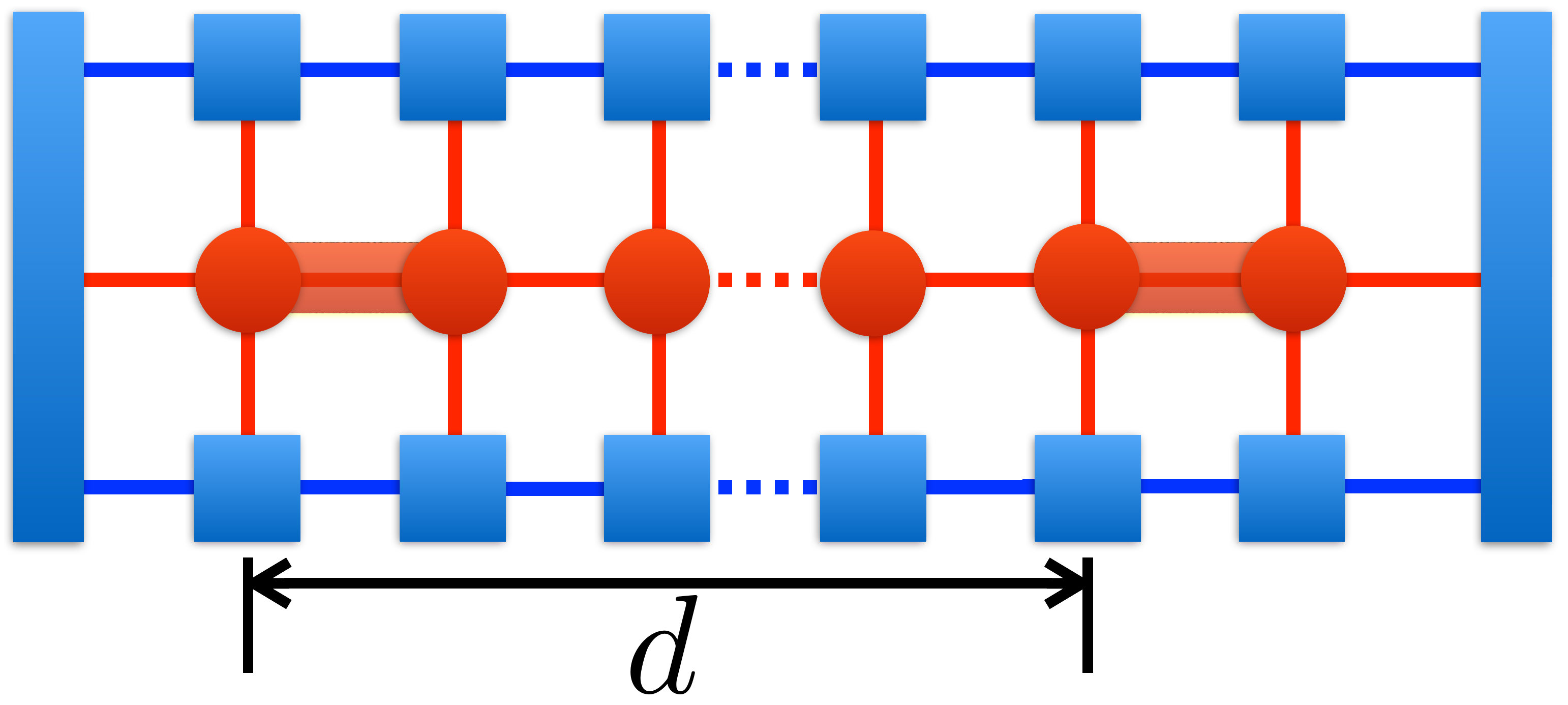}} \hfill\\
\subfloat[]{\includegraphics[width=0.55\columnwidth,angle=0]{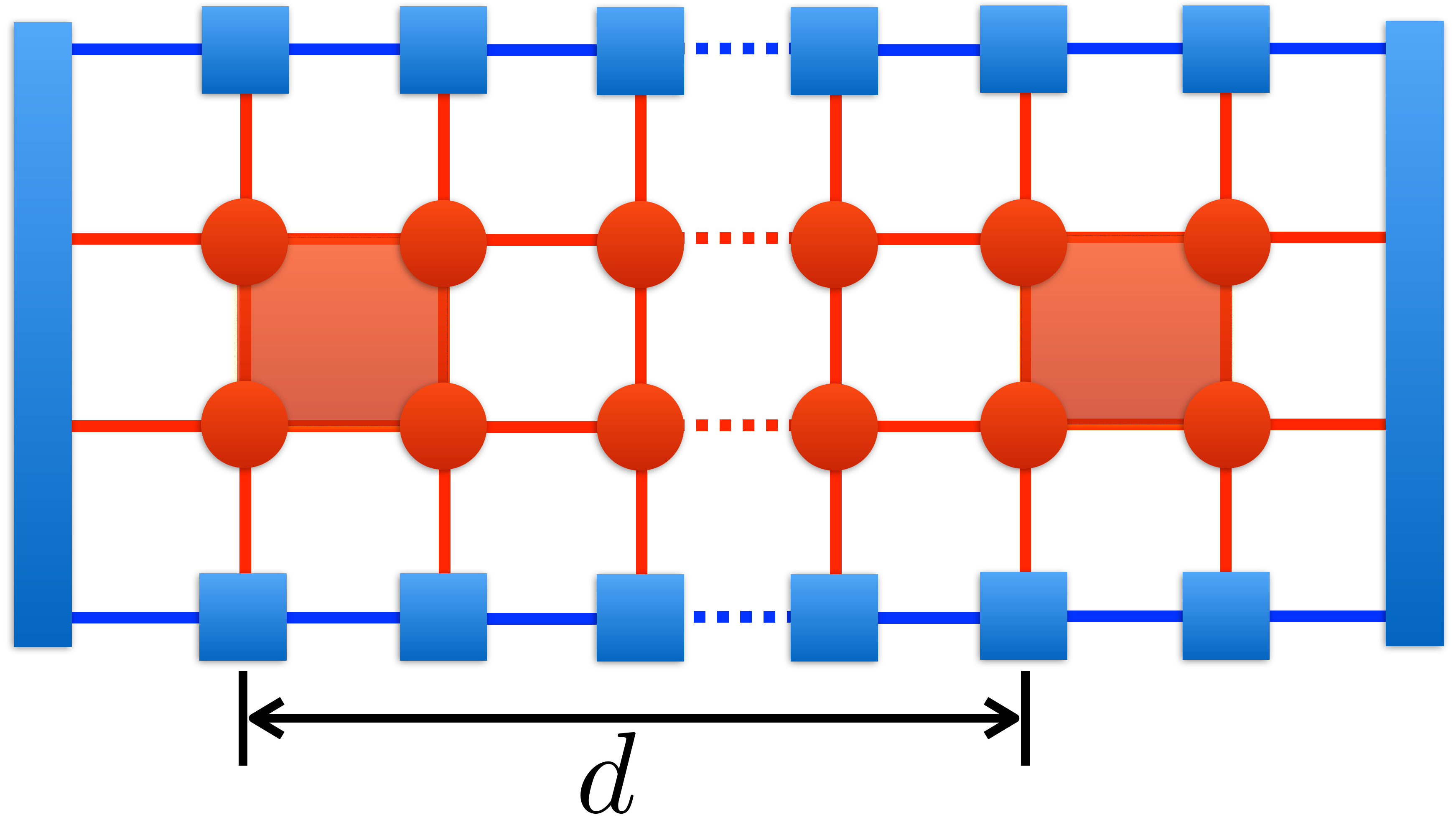}} \hfill
\caption{
The fixed point environments for strips with a single line (a) or with two lines (ladder)  (b) of $E$ tensors are obtained as 
leading eigenvectors of the transfer matrices marked by dotted boxes.
Strips to compute spin-spin (c), (longitudinal) dimer-dimer (d) and chiral-chiral (e) correlations, inserting two spin, dimer or chiral operators
within the E tensor(s) at distance $d$. }
\label{fig:CORR}
\end{figure}

Correlation between any two (local) operators can be obtained by using infinitely-long strips (running, let say, along the $x$-direction)
bounded on each side by lines of fixed-point $T$ environment tensors, as shown in Fig.~\ref{fig:CORR}(a-e).
Depending on the type of operator to be considered, a single line or two lines of $E$ tensors have to be inserted
between the two (infinitely-long) boundaries. The new fixed point environment on the left side (right side) of the left-most (right-most) operator 
is then constructed, as shown in Fig.~\ref{fig:CORR}(a,b). Two single-site, two-site or four-site  operators like
\begin{eqnarray}
 O_i^{\rm spin}&=&{\bf S}_i, \nonumber \\
 O_i^{\rm dimer}&=&{\bf S}_i\cdot {\bf S}_{i+e_x}, \\
 O_i^{\rm chiral}&=&{\bf S}_i\cdot({\bf S}_{i+e_x}\times {\bf S}_{i-e_y})+{\bf S}_i\cdot({\bf S}_{i+e_x-e_y}\times {\bf S}_{i-e_y}), \nonumber\\
+  {\bf S}_i\cdot&({\bf S}&_{i+e_x}\times {\bf S}_{i+e_x-e_y})+{\bf S}_{i+e_x}\cdot({\bf S}_{i+e_x-e_y}\times {\bf S}_{i-e_y}), \nonumber
\end{eqnarray}
where $e_x$ ($e_y$) is the unit vector along (perpendicular) to the strip are then inserted at a distance $d$, as shown in \hbox{Fig.~\ref{fig:CORR}(c-e)}. 
The corresponding correlation functions can then be computed by applying $d-1$ or $d-2$ times the TM between the two operators.  
Note that, when the local operator has a finite expectation value, the connected 
correlation function is computed, i.e. making the replacements 
$O_i^{\rm dimer}\rightarrow O_i^{\rm dimer} - \big<O_i^{\rm dimer}\big>$ and 
$O_i^{\rm chiral}\rightarrow O_i^{\rm chiral} - \big<O_i^{\rm chiral}\big>$.

\begin{figure}[htbp]
\centering
\includegraphics[width=0.99\columnwidth,angle=0]{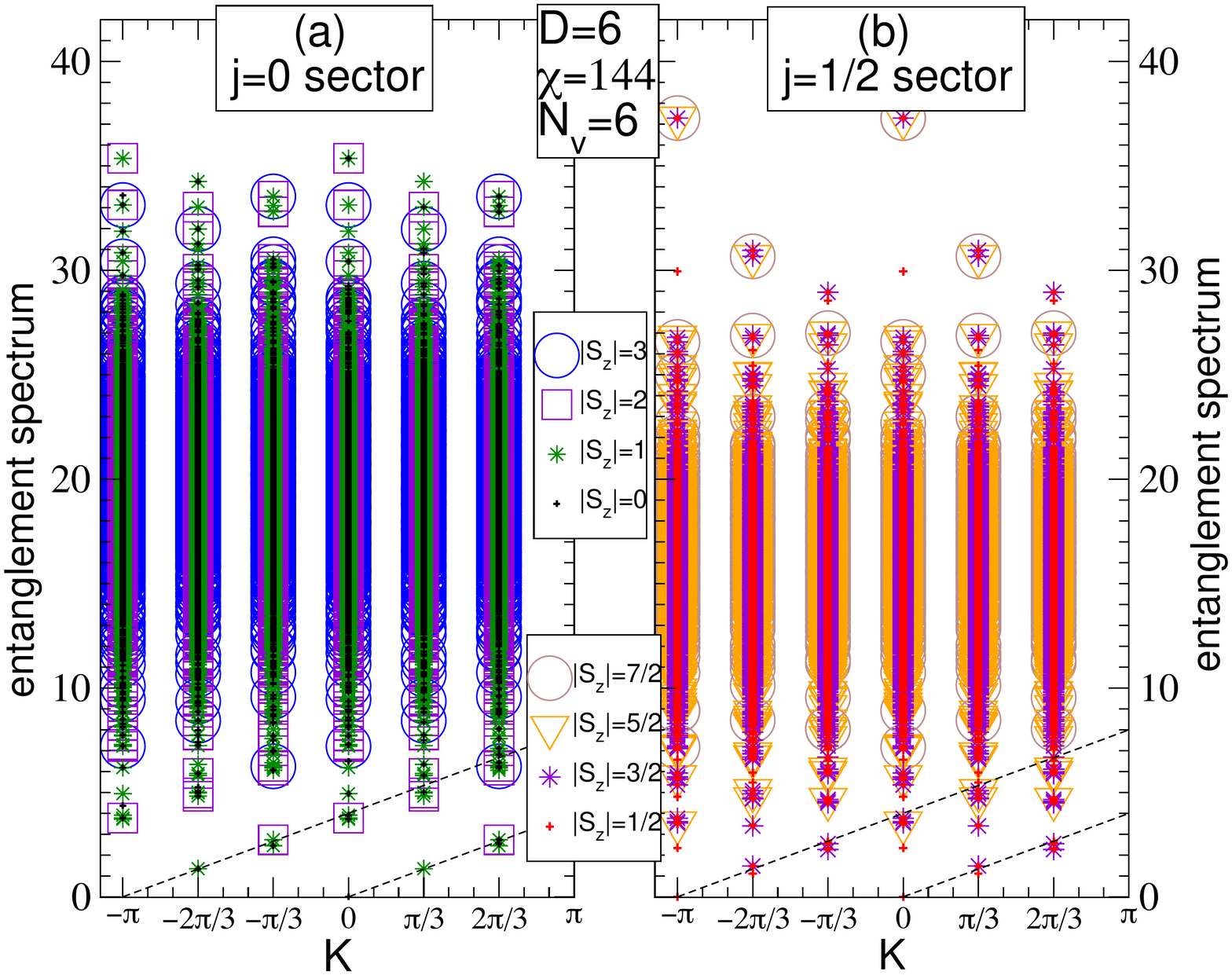}
	\caption{
Complete entanglement spectrum of the $B_1+iB_2$ $D=6$ chiral PEPS for $N_v=6$, versus edge momentum $K$.
Different symbols correspond to different
values of  $|S_z|$, showing that the spectrum is composed of exact $SU(2)$ multiplets with integer (a) and half-integer spins (b). 
Dashed lines correspond to the low-energy chiral CFT modes. }
\label{fig:ES_chi144_ALL}
\end{figure}

\begin{figure*}[htbp]
\centering
\includegraphics[width=0.99\columnwidth,angle=0]{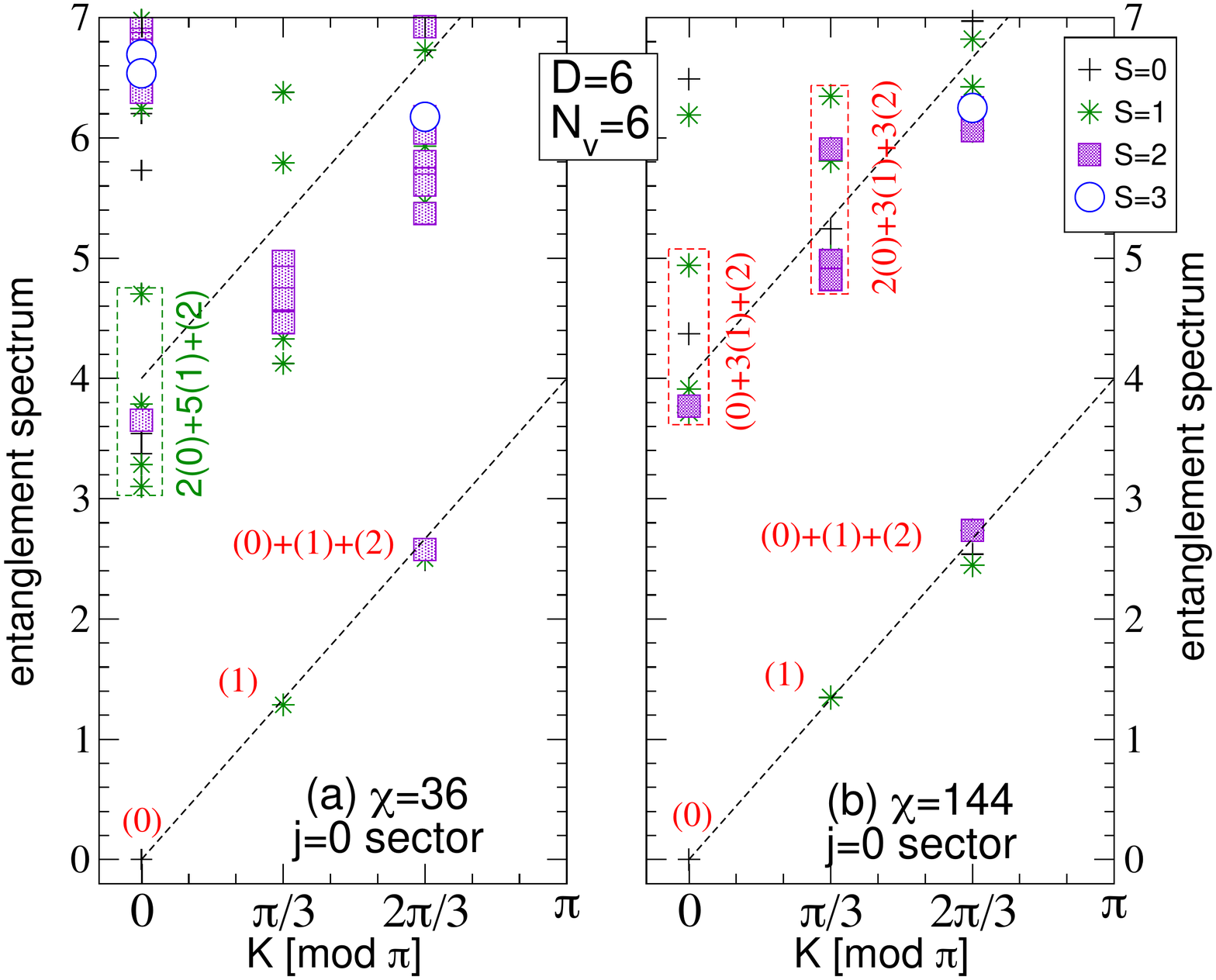}
\includegraphics[width=0.99\columnwidth,angle=0]{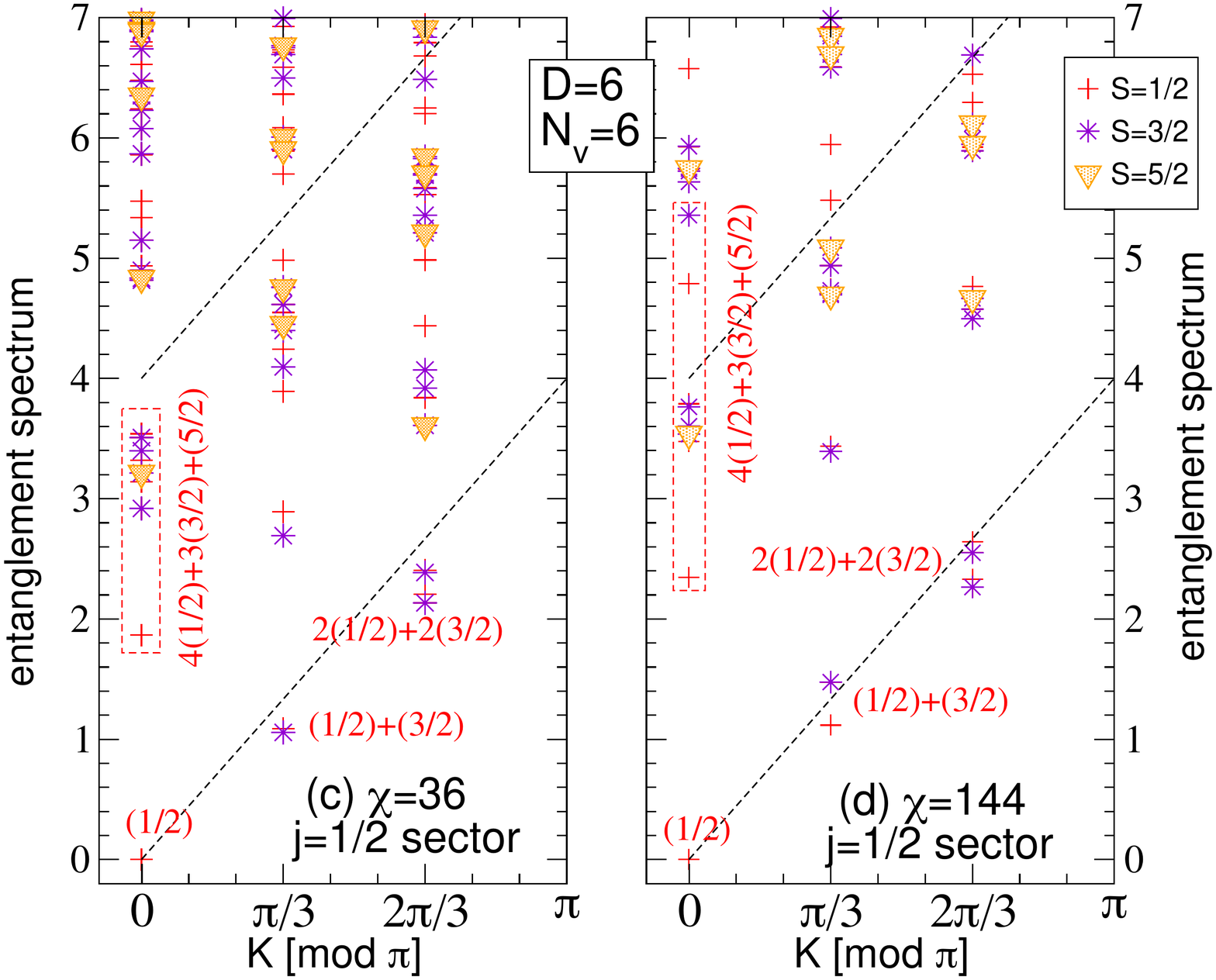}
	\caption{
Comparison between entanglement spectra of the $B_1+iB_2$ $D=6$ chiral PEPS (for $N_v=6$) computed using the iPEPS $T$ tensor obtained either with $\chi=D^2=36$ (a,c) or with $\chi=4D^2=144$ (b,d). }
\label{fig:ES_comparison}
\end{figure*}

\section{Additional data on the entanglement spectrum for $N_v=6$}
\label{app2}

For completeness we provide in Fig.~\ref{fig:ES_chi144_ALL} the full entanglement spectra in the $j=0$ and $j=1/2$ topological sectors,
computed from a ring of $N_v$ $T$ environment tensors at $\chi=144$.
Note that in the gauge chosen to write down the PEPS (convenient for CTMRG), the generator of the spin-SU(2) group are invariant 
under even translation and since $S_z\rightarrow -Sz$ under odd translations. As a consequence $S_z$ is defined up to a sign. 
This also implies that the $2S+1$ states of the SU(2) spin-$S$ multiplets (labelled by $|S_z|$ instead of $S_z$) are split between 
momentum $k$ and momentum $k+\pi$, as can be checked directly in Fig.~\ref{fig:ES_chi144_ALL}. 
Note that the two dashed lines at low energy correspond, in fact, to a unique chiral mode as it becomes clear by plotting the spectrum in
the reduced Brillouin zone $[0,\pi[$ (see main text). Note that the spectrum can be ``unfolded'' and plotted in the full Brillouin zone while 
still keeping
the full SU(2) multiplet structure by using a different gauge for the PEPS that does not preserve the full rotation symmetry of the local
tensor~\cite{Hackenbroich2018}.

Since the ES of the $D=6$ chiral PEPS is computed using the CTMRG fixed-point tensor $T$, a systemic finite-$\chi$ error seems inerrant to the calculation. Nevertheless, one can argue that the results should saturate once, typically,  the
correlation length $\xi_{\rm MPS}^{(1)}(\chi)$ becomes bigger than the system size $N_v$. The ES obtained for $\chi=36$ 
and $\chi=144$, for which $\xi_{\rm MPS}^{(1)}\sim 2.8$ and $\xi_{\rm MPS}^{(1)}\sim 5.8$ respectively, 
are compared in Fig.~\ref{fig:ES_comparison}(a-d). Clearly, the spectra for $\chi=36$ are not fully converged, e.g. the 4th $j=0$ Virasoro level (shown by the green box) containing spurious $S=0$ and $S=1$ multiplets. The correct SU(2)$_2$ counting is obtained for more
levels for $\chi=144$. Since $\xi_{\rm MPS}^{(1)}(\chi=144)\sim N_v$, we expect that this spectrum is already quite close from the exact
ES of the $D=6$ chiral PEPS on an infinitely-long cylinder of perimeter $N_v=6$.

\end{document}